\documentclass[longauth]{aa} 

\usepackage{graphicx}
\usepackage{subcaption}
\usepackage{mwe}
\usepackage{booktabs}
\usepackage{tabularx}
\usepackage{multirow}
\usepackage{natbib}
\usepackage{xcolor}
\usepackage{graphicx}
\usepackage{txfonts}

\defcitealias{Salvato22}{S22}

\usepackage[bookmarks=false,colorlinks=true,linkcolor=orange,citecolor=cyan,filecolor=black,urlcolor=cyan]{hyperref}
\begin{document}


   \title{Counterpart identification and classification for eRASS1 and characterisation of the AGN content} 
   

   \author{M. Salvato\thanks{mara@mpe.mpg.de}
          \inst{1,2},
J. Wolf\inst{1,2,3},
T. Dwelly\inst{1},
H. Starck\inst{1},
J. Buchner\inst{1},
R. Shirley\inst{1},
A. Merloni\inst{1},
A. Georgakakis\inst{4},
F. Balzer\inst{1},
M. Brusa\inst{5,6},
A. Rau\inst{1},
S. Freund\inst{1},
D. Lang\inst{7},
T. Liu\inst{1},
G. Lamer\inst{8},
A. Schwope\inst{8},
W. Roster\inst{1},
S. Waddell\inst{1},
M. Scialpi\inst{9,10,11},
Z. Igo\inst{1},
M. Kluge\inst{1},
F. Mannucci\inst{11},
S. Tiwari\inst{1},
D. Homan\inst{12,8},
M. Krumpe\inst{8},
A. Zenteno\inst{13},
D. Hernandez-Lang\inst{14}
J. Comparat\inst{1},
M. Fabricius\inst{1},
J. Snigula\inst{1},
D. Schlegel\inst{15},
B. A. Weaver\inst{16},
R. Zhou\inst{17},
A. Dey\inst{16},
F. Valdes\inst{16},
A. Myers\inst{18},
S. Juneau\inst{16},
H. Winkler\inst{19},
I. Marquez\inst{20},
F. di Mille\inst{21},
S. Ciroi\inst{22},
M. Schramm\inst{7},
D. A. H. Buckley\inst{23, 24, 25},
J. Brink\inst{8},
M. Gromadzki\inst{26},
J. Robrade\inst{27},
K. Nandra\inst{1},
  }
  \titlerunning{eROSITA/DR1 counterparts and properties}
    \authorrunning{Salvato et al.}
\institute{Max-Planck-Institut f\"ur extraterrestrische Physik, Giessenbachstr. 1, 85748 Garching, Germany
\and Exzellenzcluster ORIGINS, Boltzmannstr. 2, D-85748 Garching, Germany 
\and Max-Planck-Institut f\"ur Astronomie K\"onigstuhl 17. D-69117 Heidelberg Germany
\and Institute for Astronomy and Astrophysics, National Observatory of Athens, V. Paulou and I. Metaxa, 11532, Greece
\and Dipartimento di Fisica e Astronomia "Augusto Righi", Universit\`a di Bologna,  via Gobetti 93/2,  40129 Bologna, Italy
\and INAF - Osservatorio di Astrofisica e Scienza dello Spazio di Bologna, via Gobetti 93/3,  40129 Bologna, Italy
\and Perimeter Institute for Theoretical Physics, 31 Caroline St. North, Waterloo, ON N2L 2Y5, Canada
\and Leibniz-Institut für Astrophysik Potsdam (AIP), An der Sternwarte 16, 14482 Potsdam, Germany
\and University of Trento, Via Sommarive 14, I-38123 Trento, Italy
\and Universit\'a di Firenze, Dipartimento di Fisica e Astronomia, via G. Sansone 1, 50019 Sesto F.no, Firenze, Italy
\and INAF - Osservatorio Astrofisico di Arcetri, Via Largo E. Fermi 5, 50125 Firenze, Italy
\and Institute of Astronomy, University of Cambridge, Madingley Road, Cambridge, CB3 0HA, UK
\and Cerro Tololo Inter-American Observatory/NSF’s NOIRLab, Casilla 603, La Serena, Chile
\and Faculty of Physics, Ludwig-Maximilians-Universit\"at M\"unchen,
Scheinerstr. 1, 81679 Munich, Germany
\and Lawrence Berkeley National Laboratory, 1 Cyclotron Road, Berkeley, CA 94720, USA
\and NSF National Optical/Infrared Research Laboratory, 950 N. Cherry Ave, Tucson, AZ 85719, USA
\and Lawrence Berkeley National Laboratory, 1 Cyclotron Road, Berkeley, CA 94720, USA
\and Department of Physics \& Astronomy, University of Wyoming, 1000 E. University, Department 3905, Laramie, WY 8207, USA
\and Department of Physics, University of Johannesburg, South Africa.
\and Instituto de Astrofísica de Andalucia (IAA-CSIC), Glorieta de la astronomia S/N 18008 Granada, Spain
\and Las Campanas Observatory - Carnegie Institution for Science  Av. Raul Bitran 1200, La Serena, Chile
\and Universit\'a di Padova, Dipartimento di Fisica e Astronomia Galileo Galilei, Via Francesco Marzolo, 8, 35131 Padova, Italy
\and South African Astronomical Observatory, PO Box 9, Observatory, Cape Town 7935, South Africa
\and Department of Astronomy, University of Cape Town, Private Bag X3, Rondebosch 7701, South Africa
\and Department of Physics, University of the Free State, PO Box 339, Bloemfontein 9300, South Africa
\and Astronomical Observatory, University of Warsaw, Al. Ujazdowskie 4, 00-478 Warszawa, Poland
\and University of Hamburg, Gojenbergsweg 112,
21029 Hamburg, Germany
}
\date{Received September XX, 2020; accepted XX, 2020} 
\abstract  
   {
   Accurately accounting for the AGN phase in galaxy evolution requires a large, clean AGN sample. This is now possible with SRG/eROSITA, which completed its first all-sky X-ray survey (eRASS1) on June 12, 2020. The public Data Release 1 (DR1, Jan 31, 2024) includes 930,203 sources from the Western Galactic Hemisphere.
   }
  {The data enable the selection of a large AGN sample and the discovery of rare sources. However, scientific return depends on accurate characterisation of the X-ray emitters, requiring high-quality multiwavelength data. This paper presents the identification and classification of optical and infrared counterparts to eRASS1 sources.
  }
   {Counterparts to eRASS1 X-ray point sources were identified using Gaia DR3, CatWISE2020, and Legacy Survey DR10 (LS10) with the Bayesian NWAY algorithm and trained priors. Sources were classified as Galactic or extragalactic via a Machine Learning model combining optical/IR and X-ray properties, trained on a reference sample. For extragalactic LS10 sources, photometric redshifts were computed using {\sc Circlez}.
   }
   {Within the LS10 footprint, all 656,614 eROSITA/DR1 sources have at least one possible optical counterpart; $\sim$570,000 are extragalactic and likely AGN. Half are new detections compared to AllWISE, Gaia, and Quaia AGN catalogues. Gaia and CatWISE2020 counterparts are less reliable, due to the survey's shallowness and the limited amount of features available to assess the probability of being an X-ray emitter. In the Galactic Plane, where the overdensity of stellar sources also increases the chance of associations, using conservative reliability cuts, we identify approximately 18,000 Gaia and 55,000 CatWISE2020 extragalactic sources.}
   {We release three high-quality counterpart catalogues — plus the training and validation sets — as a benchmark for the field. These datasets have many applications, but in particular empower researchers to build AGN samples tailored for completeness and purity, accelerating the hunt for the Universe’s most energetic engines.}  

   \keywords{quasars: individual --
                Galaxies: high-redshift --
                X-rays: galaxies
               }

   \maketitle

\section{Introduction}
\label{Intro}

Since the work of the Nuker team\footnote{\url{https://en.wikipedia.org/wiki/Nuker_Team}}, and in particular  following the paper of \citet{Magorrian1998}, it has become widely accepted that galaxies hosting an actively accreting supermassive black hole (SMBH) in their center are not exceptions in the extragalactic population, but rather the norm.

This recognition introduced the challenge  of incorporating the Active Galactic Nuclei (AGN) phase in galaxy evolution models, and highlighted the need to understand how the evolution of galaxies and their central SMBH  are connected, and how the two influence each other.
This interconnection is reflected in a series of more or less tight scaling relations \citep[e.g.,][]{Gebhardt2000, Ferrarese2000}, and is often attributed to AGN feedback. Through various mechanisms, this feedback can regulate, or even suppress, star formation, thereby influencing the growth of both the galaxy and its central black hole \citep[e.g., see review of][and references therein]{Harrison2024}. Among others, \citet{Kormendy2013a, Kormendy2013b} provided comprehensive reviews of the relationship between SMBHs and their host galaxies.  Their studies, however, challenge a simplistic interpretation, showing that the properties of the SMBHs correlate differently with various galaxy components, pointing to a complex evolutionary interplay.

Completeness (i.e., how fully a sample represents the intended target or population without significant omissions) and purity (the degree to which a sample is free from contaminating interlopers) are critical to investigate observationally the scaling relations between galaxies and their central black holes.
X-ray emission is primarily produced by the accretion of material onto supermassive black holes at the centers of AGN. This direct relationship allows for a more straightforward identification of AGN compared to other methods, such as optical or infrared selection, which may include contributions from star formation or other Galactic processes
\citep[e.g.,][]{Eckart2010}.
However, the method is biased against the most obscured sources, that, in contrast, are more readily identified through infrared (IR) selection criteria, which generally offer higher completeness but lower purity \citep[e.g.,][]{Hickox2018}, i.e., the efficacy in isolating AGN activity from other sources of emission, minimizing contamination from unrelated phenomena.

Most  AGN samples used to date originate either from deep observations in small-area surveys \citep[e.g.,][]{Civano12, Hsu14,Ni2021} or from shallow observations covering wide areas \citep[e.g., ROSAT/2RXS][]{Boller16}. The relation between the growth and galaxy evolution is complex and multi-parametric, and therefore, large and complete samples are needed to reveal underlying covariances among the key physical parameters. 

The situation has changed thanks to the X-ray all-sky survey of the extended ROentgen Survey with an Imaging Telescope Array (eROSITA) \citep{Predehl2021} onboard the Spectrum-RG mission \citep{Sunyaev2021}.
Already within the first 6 months of the eROSITA all-sky survey \citep[eRASS1;][]{Merloni24}, nearly one million sources have been detected in the Western Galactic hemisphere, the vast majority of which are AGN. Notably, eRASS1 uncovered many rare AGN, such as luminous or high redshift objects \citep[e.g.,][]{Wolf21,Medvedev2021}, which are typically missed by narrow, deep, pencil-beam surveys. The resulting AGN sample is highly complementary to those selected at other wavelengths and offers new insights into AGN evolution and their connection to host galaxy properties. 
Nevertheless, X-ray observations alone cannot drive the AGN/galaxy co-evolution studies. Establishing reliable associations with multi-wavelength counterparts is a critical step toward enabling the detailed characterisation of these sources, e.g., distinguishing between Galactic and extragalactic origins and deriving their redshifts via spectroscopy and photometry. For all-sky surveys, this is particularly challenging, due to the varying quality of multi-band data across the sky, the complex geometry of surveys, and the relatively low spatial resolution of eROSITA (PSF Half Energy Width of $\sim$30\arcsec\, and corresponding 1$\sigma$ median positional uncertainty of $\sim$4.5\arcsec); for the fainter eRASS1 sources, positional uncertainties exceeding 15\arcsec\ are not uncommon \citep[see][]{Merloni24}. Once counterparts are identified, their classification as extragalactic is necessary, so that redshift can be acquired with spectroscopy or computed via photometric redshift \citep[see review by][]{Salvato18b}.  
Only then can well-defined AGN samples be assembled, with selection controlled for various quality parameters, enabling robust statistical and physical analyses.

This paper describes both the methodology used to assign reliable counterparts to all eROSITA sources presented in the 0.2--2.3\,keV selected Main catalogue of the public Data Release 1 (DR1) \citet{Merloni24}, the redshift determination and the subsequent definition of the AGN sample derived from these associations, including comparisons with other AGN samples.  
Papers that have already made use of the catalogues released with this paper span a large variety of topics: from the identification of unresolved clusters in the eROSITA point-source catalogue \citet{Balzer2025}, to identification of BLAZARs (H\"ammerich et al, 2025, to be submitted); from the identification of eROSITA/DR1 variable sources \citet{Boller2025} and tidal disruption events \citep{Grotova2025} to  the identification of AGN in the Euclid Q1 fields \citep{Roster2025}; from the study of the evolution of the X-ray and UV relation \citep{Chira} to the study the X-ray properties of BAL QSOs (Hiremath et al, 2025) and soft X-ray eccess \citep{Chen2025}; from the targeting of eROSITA sources with SDSS-V \citep{Kollmeier2025} to  the spectra analysis of AGN \citep{Aydar2025, Pulatova2025}  in SDSS-V/DR19 \citep[][]{SDSSDR192025}.
A complementary analysis for the hard X-ray-selected sample was provided in \citet{Waddell2024}.
In Section \ref{section:Xray}, we briefly summarise the characteristics of the eRASS1 catalogue and the supporting multiwavelength catalogues. The determination of the counterparts is detailed in Section \ref{section:counterparts}, while Section \ref{Rosat} present a basic comparison with the counterparts to ROSAT/2RXS \citep{Boller2021}.
Sections \ref{section:CTPproperties} and \ref{sec:redshifts} focus on disentangling Galactic and extragalactic sources and on the redshift computation for the extragalactic population. The definition of the final eRASS1 AGN sample, along with the comparison to other AGN samples, is presented in Section \ref{sec:eRASS1_AGN}. Section \ref{Section: DataRelease} describes the six catalogues released with this work: catalogues of eRASS1 counterparts using different optical catalogues, AGN samples in high and low Galactic latitude regions, and the training datasets used throughout the analysis.  The summary and an outlook conclude the work in Section \ref{section:conclusions}. 

Throughout the paper, we provide AB magnitudes unless stated otherwise. In order to allow direct comparison with existing works from the literature on X-ray surveys, we adopt a flat $\Lambda$CDM cosmology with $h=H_0/[100\,\mathrm{km\, s}^{-1} \mathrm{Mpc}^{-1}]=0.7$, $\Omega_M$=0.3, and $\Omega_\Lambda$=0.7.

\section{The data}
\label{section:Xray}
Below and in Table \ref{tab:summary_survey}, we summarise the key characteristics of the main eROSITA catalogue and the supporting multi-wavelength datasets used for the identification and characterisation of counterparts.

\subsection{eROSITA/eRASS1 catalog}
\citet{Merloni24} presented the Main and Hard-selected samples of X-ray sources detected with eROSITA in the first 6 months of the survey (eRASS1) over half of the sky (Western Galactic hemisphere), the so-called eROSITA-DE region. In the Main catalogue, a uniform flux limit (at 50\% completeness) of $F_{\rm 0.5-2 keV} > 5\times 10^{-14}$ erg s$^{-1}$ cm$^{-2}$ is achieved over most of the extragalactic sky, with deeper coverage towards the Ecliptic poles. To facilitate the data analysis, the sky has been split first into overlapping tiles of 3.6 by 3.6 sq. degrees, and then  into contiguous tiles of 3 by 3 sq. degrees \citep[see Figure 2 in][]{Merloni24}. From now on, we will use the latter definition.

\begin{figure}
    \centering
\includegraphics[scale=0.42]{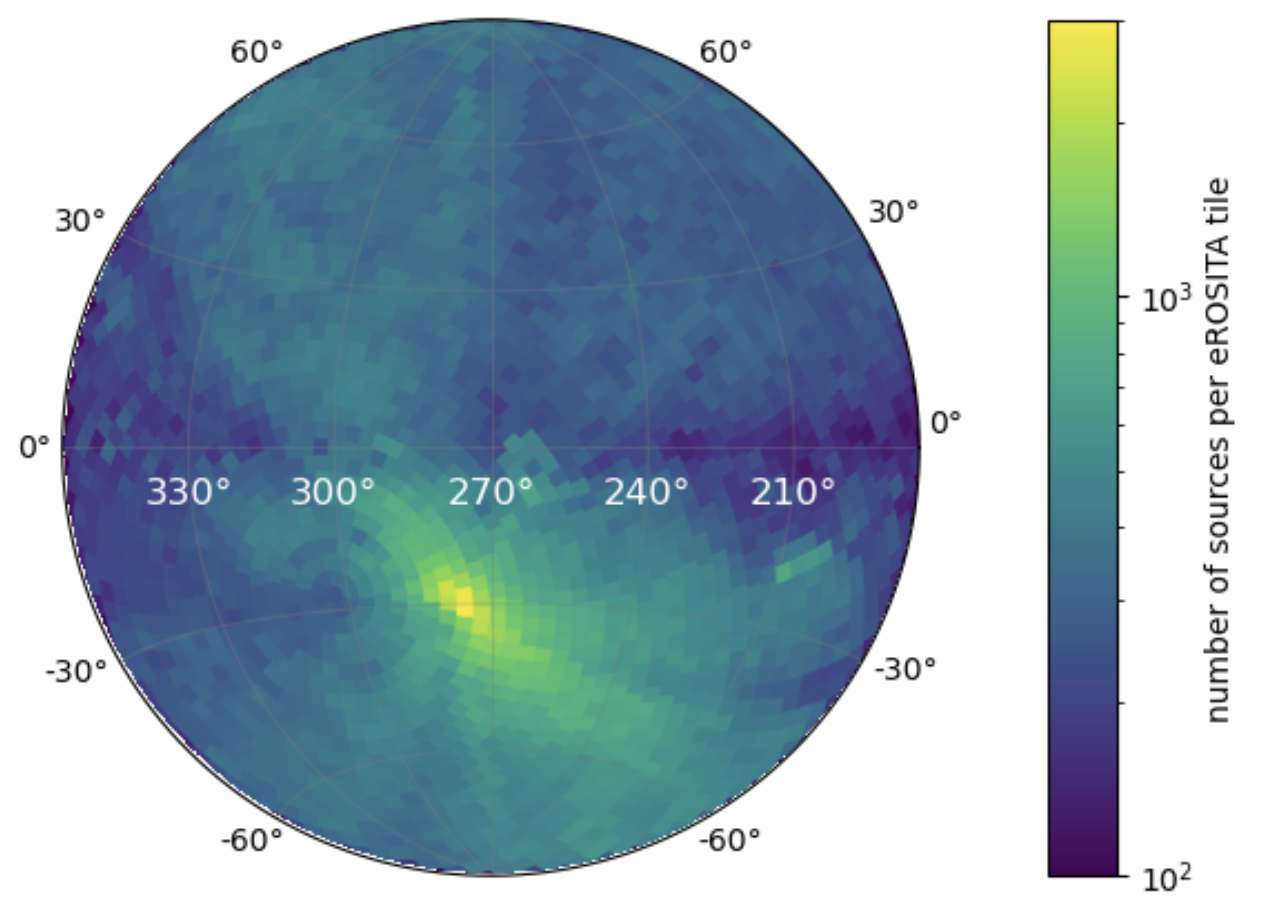}
    \caption{Number of X-ray sources detected in eRASS1 per contiguous eROSITA tile ($3\deg \times\, 3\deg$), over the entire eROSITA\_DE region in zenithal equal area (ZEA) projection. The highest density is at the South Ecliptic Pole (SEP). The map is shown in Galactic coordinates.}
    \label{fig:eRASS1density}
\end{figure}

Here, we focus on the eRASS1 Main sample, which includes  930,203 sources detected in the 0.2–2.3 keV band at various depths depending on the sky location (deeper at the South Ecliptic Pole, shallower at the ecliptic plane (see Figure~\ref{fig:eRASS1density}). 903,521 X-ray sources (97.17\%) are classified as point-like in the eROSITA image (\texttt{EXT\_LIKE}$=$0), out of which 23,648 are flagged as potentially spurious.  The catalogue also provides the detection likelihood \texttt{DET\_LIKE\_0}, measured by PSF-fitting; at the lower detection likelihood,  \texttt{DET\_LIKE\_0=6}, the probability that the source is a spurious detection is estimated to be 14\,\% \citep{Seppi2022}, decreasing to 4\,\% for sources at \texttt{DET\_LIKE\_0}$\ge$8. We will discuss the quality of the associations as a function of \texttt{DET\_LIKE\_0} in various sections of the paper.
More technical details about the characteristics of the X-ray catalogue are presented in \citet{Merloni24}. 

\subsection{Supporting optical/IR data}
\label{section:ancillary}
The reliable identification and classification of the counterparts strongly depend on the availability of ancillary data and their quality.  Most wide-area optical imaging surveys emphasise observations of the extragalactic sky and thus avoid the Galactic plane, where an excess of Galactic sources and thus source confusion make the region challenging for extragalactic studies. In general, no single optical/IR survey is sufficiently wide, deep and complete for our purposes. Therefore, we have searched for counterparts to eRASS1 sources from a small set of complementary catalogues, described below.   \\

\begin{figure*}
    \centering
    \includegraphics[width=\textwidth]{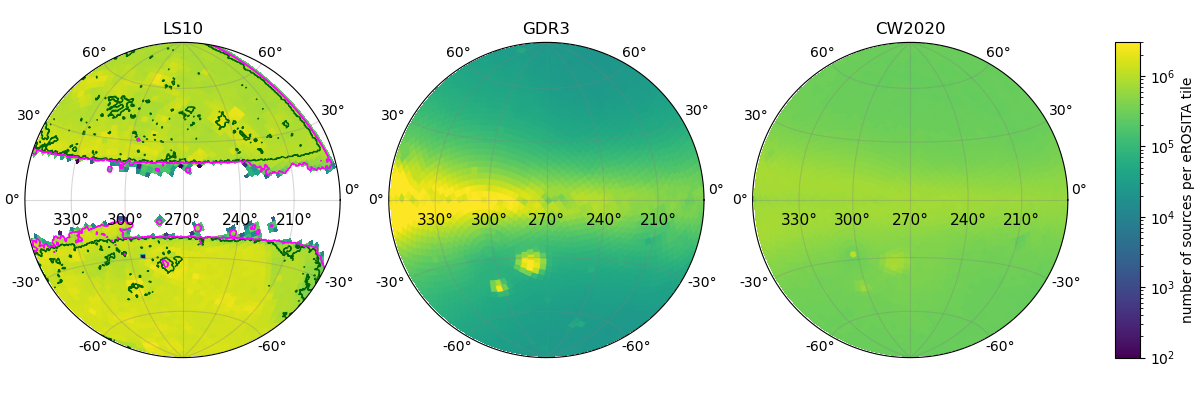}
    \caption{Source density per eROSITA sky tile  ($3\deg \times\, 3\deg$) of LS DR10 (LS10; left panel), Gaia DR3 (GDR3; middle panel) and CatWISE2020 (CW2020; right panel). On the LS10 map, the dark green (magenta) regions of InAllLS10 (InAnyLS10), indicating whether all (at least one of) the LS10 bands reach the nominal depth of the survey (see Section \ref{sec:limitations}), are overplotted.}
    \label{fig:densityfield}
\end{figure*}

\subsubsection{Data Release 10 of the Legacy Survey imaging in support of DESI} 
\label{subsec:LS10}
The Data Release 10 of the Legacy Survey imaging in support of DESI (LS10) was made public in January 2023, and while the data analysis is similar to Data Release 9 \citep[][]{Dey2019}, two significant differences characterise LS10. First, the survey covers the entire Southern extragalactic sky \citep[see Figure 1 of][]{Saxena2024}. This was possible thanks to the inclusion, among others, of DECam data taken under the DeROSITAS  program \citep[P.I.Zenteno;][]{Zenteno2025}, designed to provide ancillary data to eROSITA for cosmological studies \citep[e.g.,][]{Bulbul2024, Ghirardini2024A}, outside the area of DES \citep{DES2005}. Second, the survey includes {\it i} band data, which increases the number of detected sources\footnote{About 30\%  more sources in the test area in the COSMOS field are detected using {\it griz}, instead of {\it grz} (DESI, priv. comm.).}.

LS10 includes the photometry in the WISE near and mid-infrared bands measured at the position of the optical sources \citep{Lang14}, using, for W1 and W2, the data acquired through year 7 of NEOWISE-Reactivation\footnote{\url{https://wise2.ipac.caltech.edu/docs/release/neowise/neowise\_2021\_release\_intro.html}} and the AllWISE data for W3 and W4. For sources that are detected also by Gaia \citep[e.g.,][]{GaiaMission}, the basic astrometric and photometric Gaia DR3 columns are also included in the catalogue. The projected sky density of the LS10 catalogue per eROSITA tile is shown in the left panel of Figure~\ref{fig:densityfield}.
 
\subsubsection{Gaia DR3 (GDR3)}
\label{subsec:Gaia}
Gaia  \citep[e.g.,][]{GaiaMission} provides an all-sky catalogue with precise coordinates of optical point-like sources (i.e., stellar objects, QSOs, and compact nuclei of galaxies with a limiting magnitude of about G=20.7, in three bands (G, BP, RP). In addition, it provides reliable information on variability \citep[][]{GaiaVariability23}, proper motion, and parallax. Last but not least, the third data release \citep[GDR3;][]{GaiaDR3} also provides low-resolution spectra from the blue (BP) and red (RP) photometers \citep[][]{Montegriffo2023}. Based on spectra, photometry and astrometry, five different machine learning algorithms \citep[][]{Delchambre2023} are used (a) to divide the sources in a probabilistic manner into broad astrophysical classes (single stars, white dwarfs, binaries, galaxies and quasars); (b) to determine the redshift of the sources depending on their classification; (c) to estimate the total Galactic\footnote{Indicated as Milky Way, MW in this paper} extinction. In this way, GDR3 provides a classification for 1.592 billion sources; approximately 11.4 milion sources are classified as extragalactic, 7.8 million of which have a redshift estimation \citep[see][]{GaiaDSC32023}. The projected sky density of the GDR3 catalogue per eROSITA tile used here is shown in the middle panel of Figure~\ref{fig:densityfield}.

\subsection{CatWISE2020 (CW2020)}
\label{subsec:CW2020}
The CatWISE2020 survey \citep[CW2020;][]{Marocco2021} covers the entire sky with the WISE observations from 2010 and 2018 in W1 and W2. The catalogue is generated using {\sc{crowdsource}}  \citep[][]{Schlafly2021} and its 90\,\% completeness depth is 17.7 and 17.5 magnitude (in Vega system) for W1 and W2, respectively. As for eROSITA, also the depth of CW2020 varies substantially across the sky, being deeper at the Ecliptic poles and shallower in the Ecliptic plane. In addition, CW2020 is about 1 magnitude shallower in the Galactic plane than elsewhere due to source confusion. The catalogue includes 1.89 billion sources, and the projected sky density of the CW2020 catalogue per eROSITA tile used here is shown in the right panel of Figure~\ref{fig:densityfield}.

\begin{table*}
\centering 
\small
\caption{Summmary properties of the surveys used in this paper} 
\begin{tabular}{c|c|c|c}
\hline
& & \\
   Survey & Area coverage (square degrees)  & Depth (Approx.) &Number of sources\\
   & & \\
   \hline
   \hline
   & & \\
eROSITA/DR1 & 20,626.5  & $F_{\rm 0.5-2 keV} > 5\times 10^{-14}$ erg s$^{-1}$ cm$^{-2}$ &930,203\\
LS10 &9,582 & r=23.3 & $2.81 \times 10^9$\\
GDR3 &41,252.961 (all sky)& G=20.7&$1.59 \times 10^9$\\
CW2020 & 41,252.961 (all sky)&W1=20.4 &$1.89 \times 10^9$\\
\end{tabular}
\label{tab:summary_survey}      
\end{table*}

\section {Counterparts determination}
\label{section:counterparts}

\begin{table*}
\centering 
\small
\caption{Final list of features adopted to identify an X-ray emitter in LS10, GDR3, and CW2020, respectively.} 
\begin{tabular}{c||c||c}
\hline
& & \\
   LS10 & Gaia DR3 & CW2020 \\
   & & \\
   \hline
   \hline
   & & \\
{\it g, r, i, z, W1, W2} MW extinction corrected model fluxes & Gaia \textit{G, BP, RP} mag, corrected for MW extinction& W1 and W2 fluxes\\
Model flux errors & Gaia colors, corrected for MW extinction &W1 and W2 Flux errors\\
All possible colours  MW extinction corrected & Gaia proper motion and error &W1-W2 colour \\
Shape\_r, shape\_e1, shape\_e2, $n_{Sersic}$ &Gaia parallax and error&W1 and W2 apertures photometry\\
Gaia photometry & Gaia astrometric\_excess\_noise &proper motion and error\\
Gaia S/N &  &\\
Gaia proper motion and error & & \\
Gaia parallax and error  & & \\  
   & & \\
\hline
& & \\
Recall fraction 87.4\% & Recall fraction 75.6\%  & Recall fraction 85.1\%\\
Leakage 0.7\% & Leakage 0.4\% & Leakage 6\% \\
\end{tabular}
\label{tab:features}      
\end{table*}

\begin{figure*}
    \centering
    \includegraphics[width=0.32\linewidth]{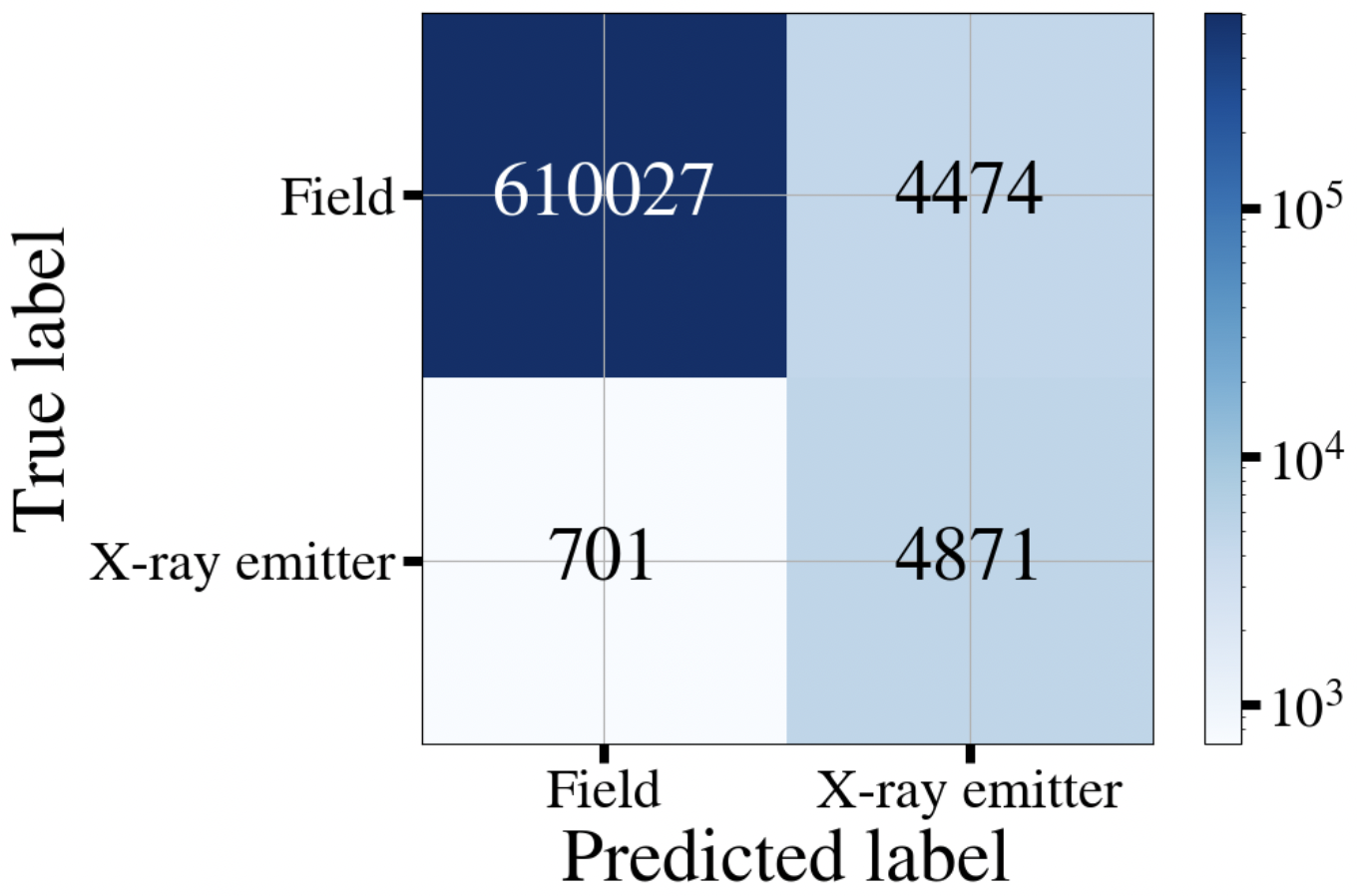}
   \includegraphics[width=0.32\linewidth]{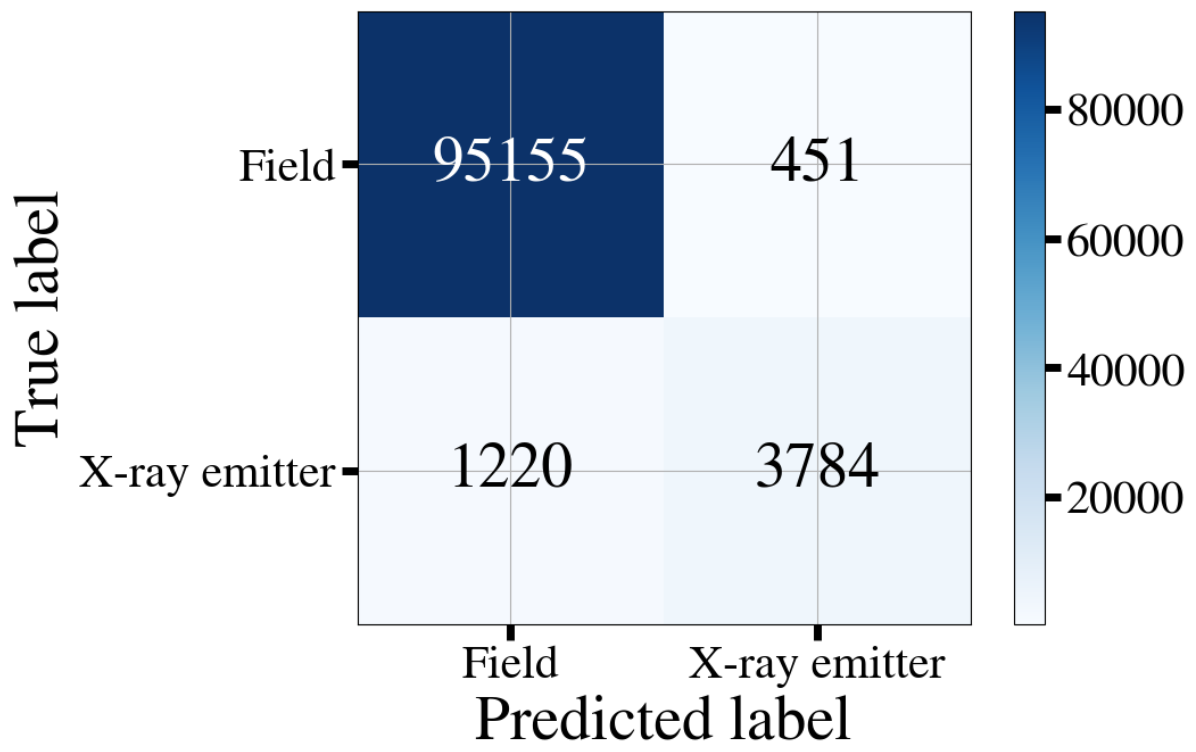}
    \includegraphics[width=0.32\linewidth]{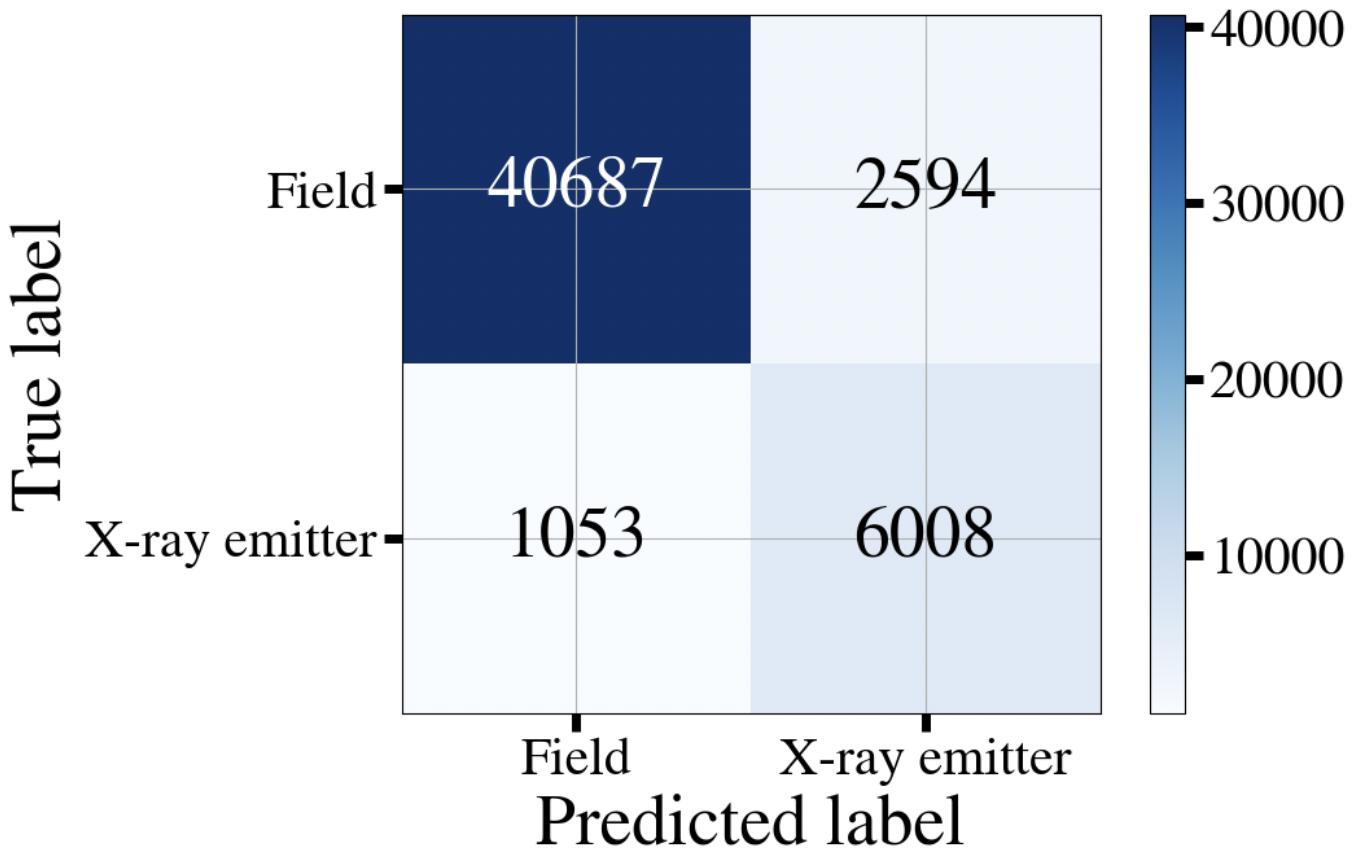}
    \caption{Confusion matrices resulting from the Random Forest classification on the validation test sets are shown for LS10 (left panel), GDR3 (middle panel), and CW2020 (right panel). The colour scale in the panels represents the number of sources per cell, with darker shades indicating higher counts. The entries along the main diagonal (top-left to bottom-right) indicate correctly classified sources. The different total number of sources across the three matrices reflects the variations in survey depth, spatial resolution, and source density among the three surveys.
   }
    \label{fig:confusionMatrix}
\end{figure*}

It is important to remember that one of the parameters that determines the reliability of the counterparts is the number density of the sources (see Figure \ref{fig:densityfield}) in the multiwavelength supporting data (the higher the source density, the larger is the chance of random associations). In turn, the number density depends on the wavelength of the survey, the depth, and the sky coordinates. This, combined with the smoothly increasing depth of eROSITA from the Galactic plane to the Ecliptic poles, challenges our ability to distinguish the correct counterparts from chance associations. In this section, we discuss how we determined the most likely counterparts to the eROSITA sources.

Given the different nature of the X-ray emission, the identification of the counterparts is made separately for sources classified as extended and point-like in the eROSITA images.
Extended X-ray sources are usually indicate the presence of clusters of galaxies \citep[][]{Kluge2024} or supernova remnants (SNR), while point sources identify stellar objects in the Milky Way or in the very nearby galaxies (Large and Small Magellanic Cloud), active black holes in galaxies (AGN, QSO) or high-redshift, etxragalactic transients of various origins \citep[e.g.,][]{Grotova2025}, and unresolved clusters \citep[][]{Balzer2025}.

There are two approaches to the determination of the counterparts. In the first case, one is interested in determining all sources of a certain type. Within the eROSITA-DE collaboration, this is the approach followed in  \citet{Kluge2024}  (focusing on the identification of clusters), H\"ammerich et al.,  (in prep; focusing on the identification of Blazars), \citet[][ focusing on the identification of  Cataclysmic variables, CVs]{Schwope2024a, Schwope2024b}, Avakyan et al., (in prep; focusing on X-ray binaries), and \citet[][focusing on the search for coronal stars]{Freund2024}. This approach is extremely reliable for defining samples of sources with high purity. However, it does not account for the possibility that the emission comes from a completely different type of source (e.g. AGN).

The second approach is to account for every type of X-ray emitter within a certain distance from the X-ray source and attempt a classification only afterwards. This is the approach adopted in the past by our group for ROSAT and XMMSlew2 \citep{Salvato18b}, in eROSITA/eFEDS \citet[][]{Salvato22, Salvato18b}, for Euclid/Q1 \citep{Roster2025} and in this paper. In all three cases, the Bayesian algorithm \texttt{NWAY} \citep{Salvato18a} was used, in combination with a prior that effectively predicts, from non-X-ray data alone, for each source the probability to be an X-ray emitter.
For the complete description of the algorithm, readers are referred to section B4 in the Appendix of \citet{Salvato18b}, while the principles and the method for constructing the prior are described in detail in \citet{Salvato22} (S22 hereafter). We provide here a summary of the concept, followed by its application in order to determine the eRASS1 counterparts in LS10, GDR3, and CW2020. \\

\subsection{eRASS1 counterparts using NWAY}\label{subsec:NWAY}
We use the Bayesian cross-matching algorithm \texttt{NWAY} to generate counterpart catalogues for the eRASS1 point sources in three supporting surveys independently: LS10, GDR3, and CW2020. For each source in the primary eRASS1 catalogue  \texttt{NWAY} ranks all the sources in the supporting catalogues within 60\arcsec\ from the X-ray position, combining the probabilities a) based on the astrometric configuration (i.e., the positional error-normalised distances separating X-ray centroid and counterpart candidates and accounting for local source density) and b) on the candidates' non-astrometric properties such as photometry, colours, morphology, or parallax. The two components of the final probability are then combined in the probability \texttt{p\_i}, following Section 3.1 of \citetalias{Salvato22}. 

The non-astrometric information is condensed in a random forest model, assigning a probability of being X-ray detected to each counterpart candidate. Following \citetalias{Salvato22}, the model was constructed using as training samples  a) secure counterparts (regardless of whether the sources are compact objects, stars, AGN, or QSO) from 4XMM DR13 \citep{Webb2020} and  {\it Chandra} CSC2.0\footnote{\texttt{primary\_source} table via the database query web
API (\url{https://cxc.cfa.harvard.edu/csc/cli/index.html})} and b) field sources within 60\arcsec\ of the X-ray position, after removing the correct counterparts. The construction of the training samples is described in Appendix \ref{appendix:4XMMBright} and released with this paper (see Section  \ref{Section: DataRelease}).

\subsubsection{Strength of the non-astrometric information} \label{subsub:priors}
As described in Appendix \ref{app: ctp_to_training}, for the training of the random forest
priors, we have extracted all LS10, GDR3, and CW2020 catalogue sources within 60\arcsec\ of the X-ray centroids of the 4XMM and Chandra training samples, thereby including the real multiwavelength counterparts as well as non-associated ``field'' sources. For each catalogue, a random forest (\texttt{sklearn} implementation, \citealt{Pedregosa11}) we formally defined a sample of true X-ray counterparts (astrometrically best match counterpart to the 4XMM or Chandra source with a maximal separation of $<1''$) and field sources (separation $>1''$)  as the input training sample. For each training sample, a subset of columns was extracted from the catalogues (e.g., fluxes, colours, signal-to-noise, etc.). Using these features, we train the model as a classifier to correctly identify true counterparts and field sources, also accounting for the missing features for some of the sources. The final set of features, listed in Table \ref{tab:features}, was refined during model tuning by sorting for feature importance (\texttt{feature\_importances\_}). The models are optimised for precision and recall fraction\footnote{In this context, a true positive is a correctly identified X-ray emitter, a true negative is a mislabelled X-ray emitter.}, defined as true positives / (true positives + false negatives), and fractional leakage, defined as false positives / (false positives + true negatives). Our final classifiers on a test sample reach a minimum recall fraction of 75.6\,\% and a maximum fractional leakage of 0.7\,\%, as shown in Figure \ref{fig:confusionMatrix}. These trained models are then used as a pre-processing step of the {\sc{NWAY}} run by predicting the probability of each candidate counterpart to be an X-ray emitting source. Note that while the recall leakage  fractions presented in Figure~\ref{fig:confusionMatrix} are based on discrete labels (threshold set at 0.5), we feed into NWAY a continuous variable.

Not surprisingly, the survey that provides the more reliable prior is LS10 (with a recall fraction of 87.4\,\%), due to the depth and resolution of the data and the number of features that can be used for separating X-ray and non-X-ray emitting objects, thus essentially able to parametrise the SED of the sources.  This is illustrated in Fig.~\ref{fig:confusionMatrix}, where the confusion matrices resulting from the random forest prediction on a validation test set using the final set of features from LS10, GDR3 and CW2020 are reported. CW2020 has almost the same recall fraction of LS10, but the number of false identifications is ten times higher, probably due to the limited number of features available.

It is important to note that the number of sources available in each training sample is different, with the number being the highest in LS10, making the associations more reliable under any aspect.

\begin{figure*}
    \centering
    \includegraphics[width=\linewidth]{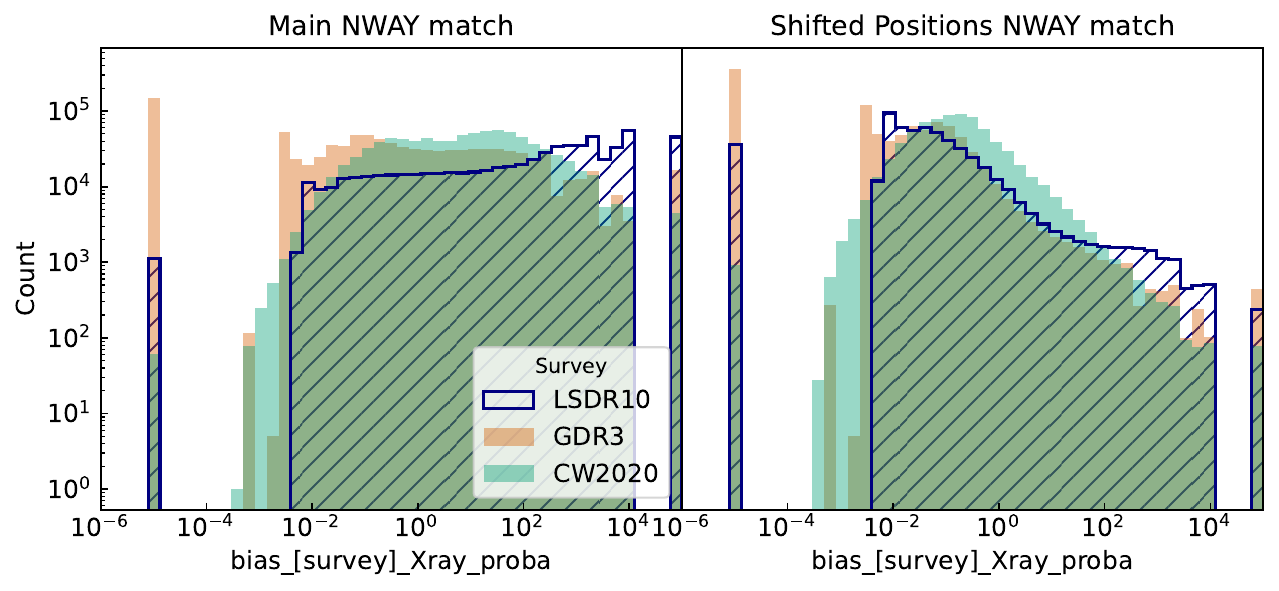}
    \caption{Histogram distribution of the probability weighting (i.e., bias), introduced by the priors for the LS10, GDR3 and CW2020 counterparts to eRASS1. The left panel shows the distribution of the bias for the actual eRASS1 counterparts, while the right panel displays the distribution for counterparts to the eRASS1 shifted positions. A value of 1 indicates no change in the probability of being the right counterpart  i.e., the probability is based solely on the distance to the X-ray position, positional uncertainties, and source number densities. Values above(below) 1 indicates that the prior has degraded(reinforced) the probability. Almost 50\% of the counterparts from GDR3 got degraded, after considering the prior. More details in the main text (Section \ref{subsub:astrometry_prior}).}
     \label{fig:bias_weight}
\end{figure*}

\subsubsection{Combining astrometry and priors}
\label{subsub:astrometry_prior}
One of the main characteristics of \texttt{NWAY} is that it is based on Bayesian statistics, allowing one to  combine via multiplication the probabilities of a source to be the correct counterpart, determined in independent ways.
So, given the coordinates of an eRASS1 source, for all the sources of a survey within 60\arcsec\ from those coordinates,  the probability of being the counterpart on the basis of astrometry is multiplied by the probability of being an X-ray emitter (assuming the corresponding random forest model). The impact that the prior has on the final probability to be the correct counterpart is expressed in the quantity \texttt{bias\_Xray\_proba} \citep[see Appendix B.6 in ][]{Salvato18a}. A value of 1 indicates no change in the probability of being the right counterpart,  i.e., the probability is based solely on the distance to the X-ray position, positional uncertainties, and source number densities. Values below 1 indicate that the prior has degraded the probability, while values above 1 indicate that the prior has reinforced the association. The left panel of Figure \ref{fig:bias_weight} shows this impact (i.e., weight)  on the counterparts from the three surveys. The sources with bias above 1 are 78\%, 55\%  and 70\% for LS10, GDR3, and CW2020, respectively. While this indicates that for LS10 and CW2020, a large fraction of the counterparts would have been correctly selected  from geometry, this is not the case for GDR3. For the latter, almost half of the GDR3 sources had a high probability of being the right counterparts based on distance from the X-ray position, but did not have the features typical of an X-ray emitter, indicating that they were just chance associations. The weighting has an ever stronger impact for the randomised eRASS1 catalogue (right panel of Figure \ref{fig:bias_weight}), where in all three surveys, the weighting is below 1 for most sources.

The final probability provided by {\sc{NWAY}} is encapsulated in two values, \texttt{p\_any}, and \texttt{p\_i}. For each X-ray source in the eRASS1 catalogue, the \texttt{p\_any} value gives the probability that at least one of the sources in the supporting multiwavelength catalogues is the correct counterpart. At the same time, for each source in the supporting catalogue within 60\arcsec\ from the eRASS1 detection, \texttt{p\_i} provides the relative probability of the object being the correct counterpart\footnote{For the analytical demonstration of the connection between \texttt{p\_any} and \texttt{p\_i} see the Appendix in \citet{Salvato18b}.}. 
 
The source in the supporting catalogues with the higher \texttt{p\_i} is assigned as the counterpart.
With this paper, we will provide catalogues of counterparts using the three surveys separately (see Section \ref{Section: DataRelease}) but recommend that the user uses LS10 when possible.

\begin{figure*}
    \centering
    
    \includegraphics[scale=0.6]{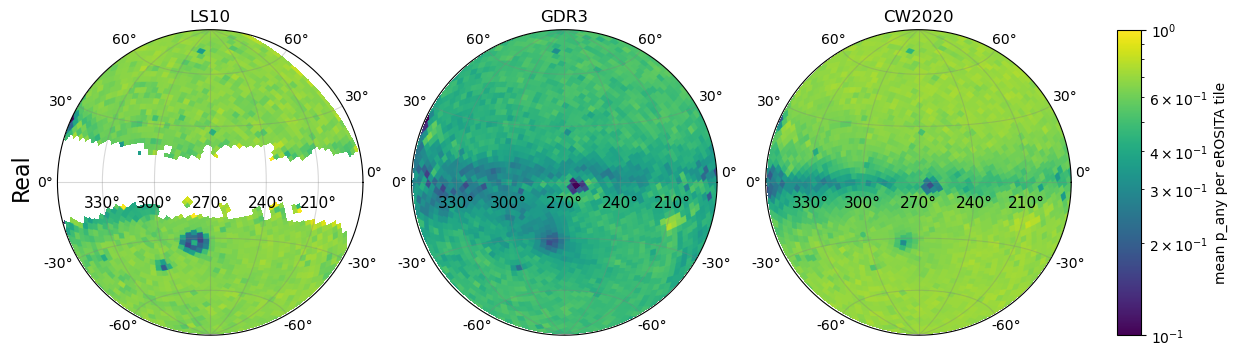}
    \includegraphics[scale=0.6]{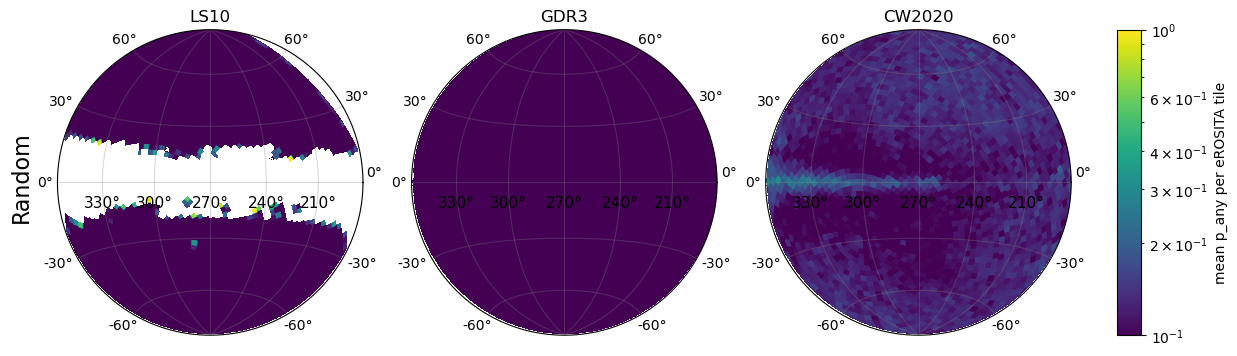}
    \caption{Mean \texttt{p\_any} distribution per eROSITA tile for real (top) and  random (bottom row) eROSITA coordinates, using ancillary data from LS10 (left),  GDR3 (middle) and CW2020 (right). The colour scale is consistent across all panels to facilitate comparison. While the \texttt{p\_any} values for the real sources are generally higher than those for the random positions, they can approach similar levels in regions of high source density, indicating an increased risk of chance associations (see the main text for further discussions).}
    \label{fig:gapa}
\end{figure*}

\subsubsection{Quantifying chance associations} \label{subsub:chance}

In \citetalias{Salvato22}, the procedure for computing the counterparts with \texttt{NWAY} in combination with machine learning for determining the prior was introduced for the first time. In the same paper, the authors quantified the \texttt{NWAY} ability to recover the correct counterpart of an X-ray source, by considering a sample of about 3500 {\it Chandra} sources\footnote{With their very small positional error, the counterparts are easily identified.}, and modifying the original very small positional error by assigning an eROSITA positional error randomly sampling from the astrometric uncertainties listed in the core eFEDS source catalogue \citep[see for details Section 4.1 of][]{Salvato18b}. It was demonstrated there that the correct counterpart was recovered in 94\,\% of the cases.

However, in that experiment, the distribution of \texttt{p\_any} in the case of chance associations was not quantified. Only knowing that, the user can decide which \texttt{p\_any} threshold and control of completeness and purity of a sample.

 In order to assess at any given \texttt{p\_any} the fraction of possible spurious associations, the coordinates of the X-ray sources have been randomised\footnote{We preserved the original position uncertainties}, and the sources with new coordinates falling within 60\arcsec\ from a known (real) eRASS1 detection, have been removed. More precisely, the sources have been shifted by 3’ in R.A. and DEC. In this way, most sources remain within the original eROSITA tile, so that the densities of primary and secondary catalogues are preserved and the probabilities can be compared.

Then, {\sc{NWAY}} is re-run on this eRASS1 random catalogue with the same settings defined for the real sample for the three supporting catalogues and the same priors. This gives us the distribution of \texttt{p\_any} values for the association of the LS10, GDR3 and CW2020 sources to a random position. Given that the new positions are random and do not fall on existing X-ray real positions, the \texttt{p\_any}should remain very small. This is because the random forest model decreases \texttt{p\_i} probability of being an X-ray emitter for the sources that could be considered counterparts from pure astrometric considerations. Figure \ref{fig:bias_weight} shows the weight of the prior in the selection of the counterparts for the real eRASS1 sources and for the randoms. If the probability of the association is unaffected by the prior, the bias will be 1, while the bias will move close to zero or to large positive values   if the prior decreases or increases the probability, respectively. The large majority of the counterparts in the randoms have a bias close to zero, indicating that the probability assigned on the basis of closeness to the X-ray coordinates where heavily rescaled when accounting for the typical physical properties of an X-ray emitter.

Figure \ref{fig:gapa} provides a further visualisation of the difference between the counterparts associated with the real and random eRASS1 sources. There, the mean \texttt{p\_any} per eROSITA tile is reported for each ancillary catalogue, both for the real eRASS1 sources (top) and for the random (bottom). Overall, the \texttt{p\_any} values are typically much higher for the eRASS1 than for the randoms, with the regions of the Magellanic Clouds and the Galactic plane having the lower values. However, the typical \texttt{p\_any} in the randoms is much lower, so that in general also in these two regions the risk of chance association is limited, with the exception of the Galactic Plane for the CW2020 (see lower right panel of Figure \ref{fig:gapa}).

\subsubsection{Completeness and purity of {\sc{NWAY}} associations} \label{subsub:compur}

\begin{figure}
    \centering
    \includegraphics[width=\columnwidth]{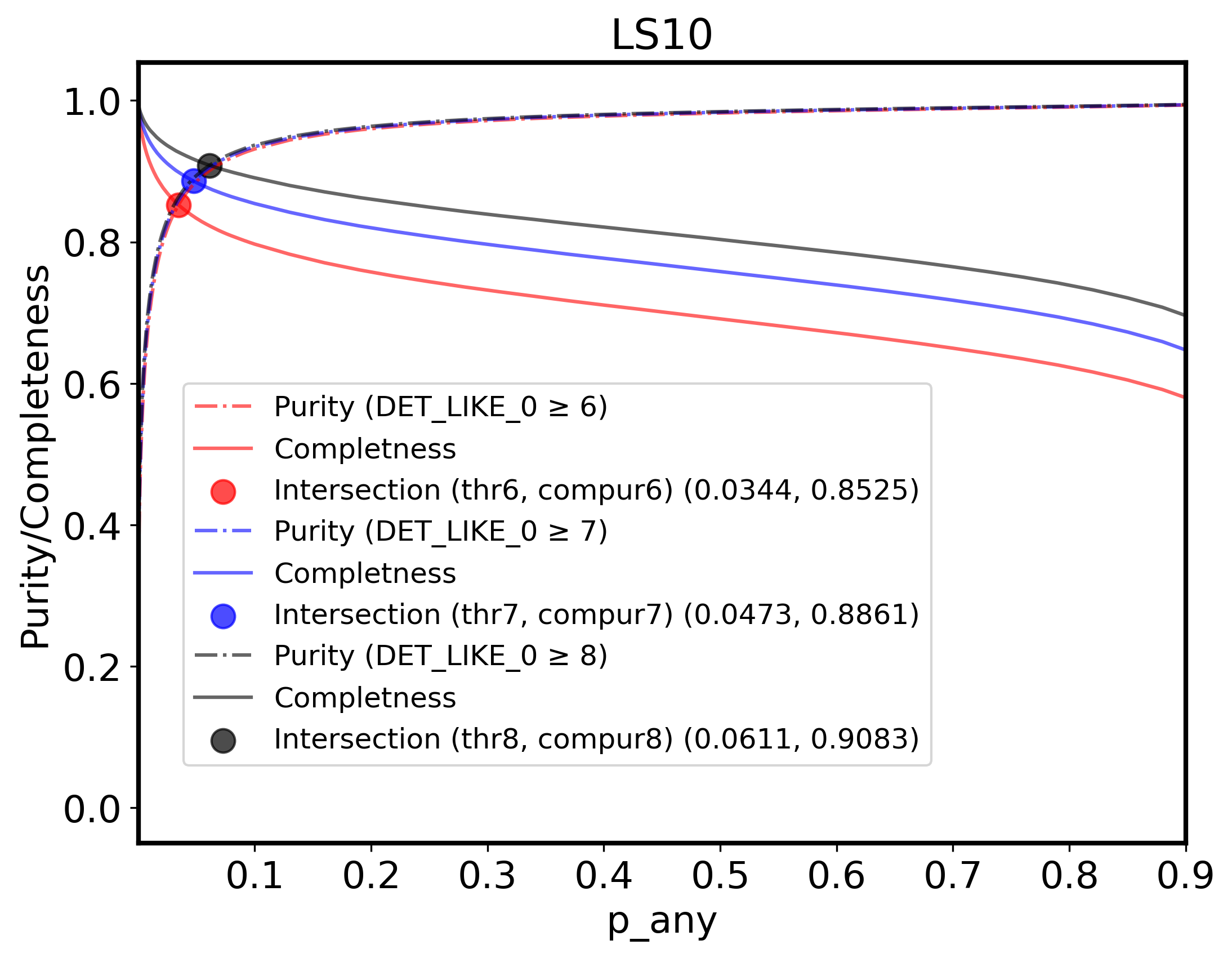}
\includegraphics[width=\columnwidth]{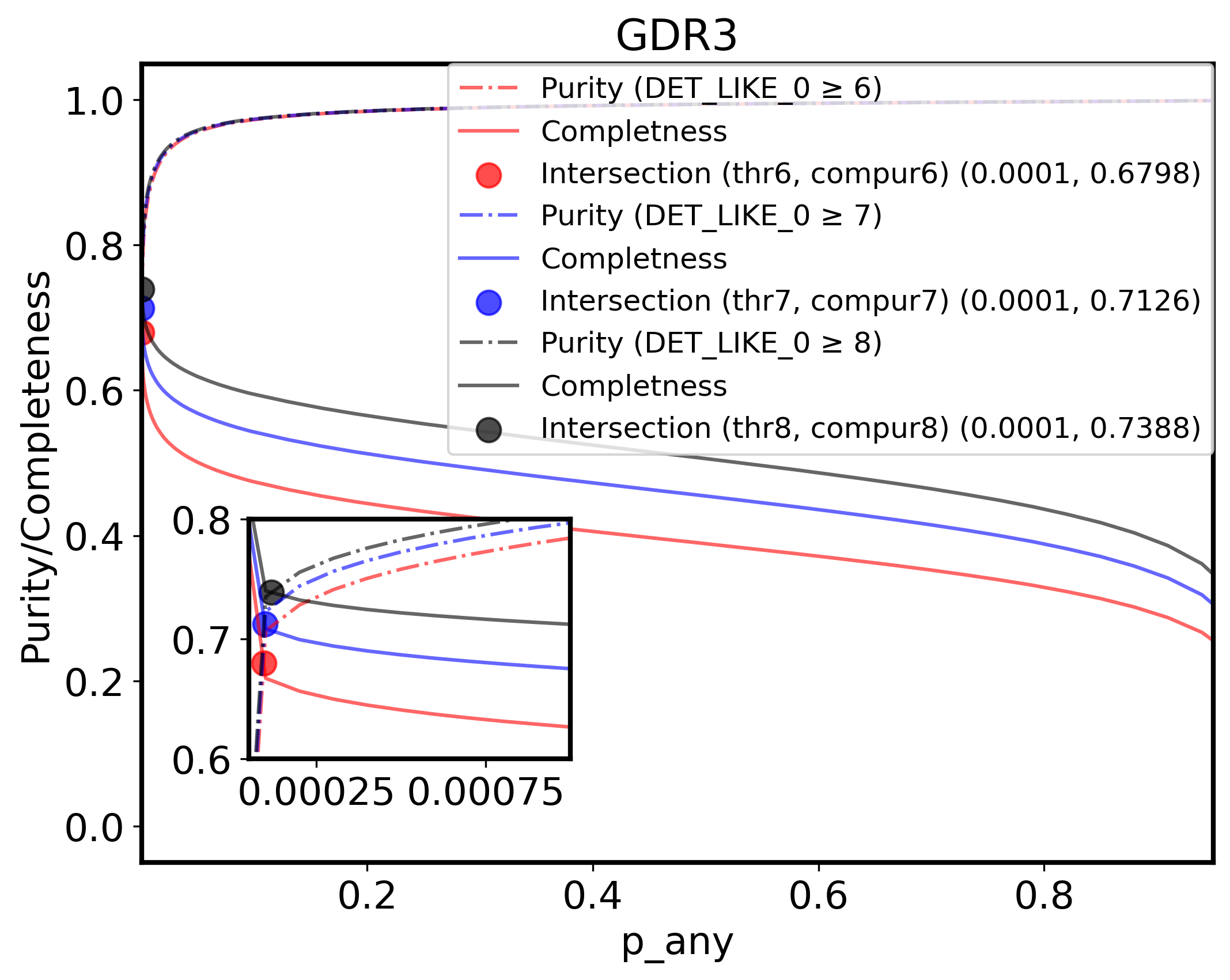}
\includegraphics[width=\columnwidth]{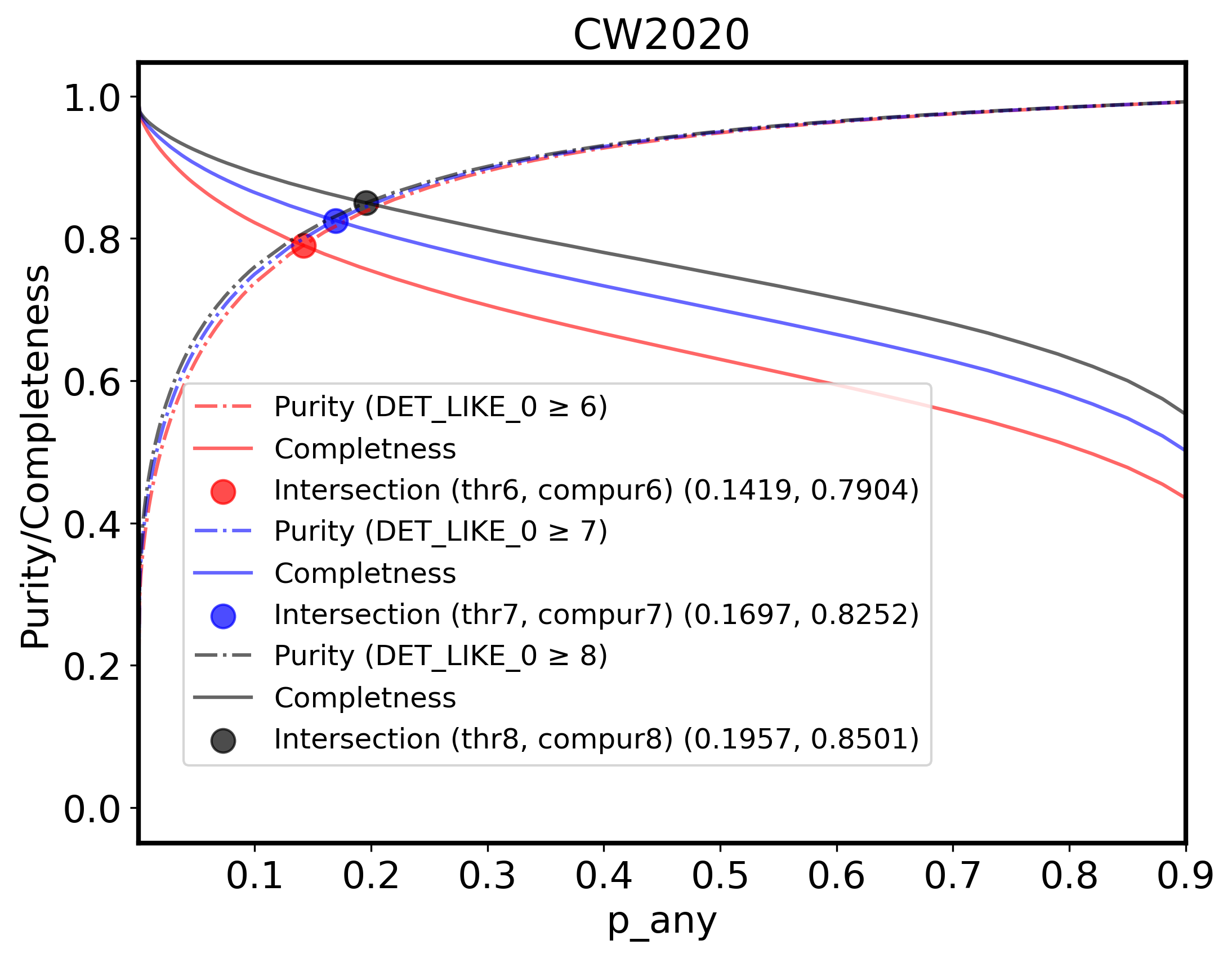}
    \caption{Mean purity and completeness as  a function of \texttt{p\_any} for LS10, GDR3 and CW2020, averaged over their respective surveys' footprints (i.e, considering eROSITA sources outside the Galactic Plane in the case of LS10)}. Different colours indicate different detection likelihood thresholds (\texttt{DET\_LIKE\_0}) for the eROSITA sources. The catalogues include the \texttt{p\_any} thresholds and the completeness/purity intersection averaged \texttt{compur}[6,7,8] computed per eROSITA tile as a function of \texttt{DET\_LIKE\_0}.
    \label{fig:Meanpurcomp}
\end{figure}

For a given ancillary catalogue, a threshold on the \texttt{p\_any} value will determine the completeness (the fraction of real eRASS1 sources that have \texttt{p\_any} above any given value) and purity (1 - the fraction of sources that have \texttt{p\_any} above any given value in the corresponding random catalogue). The threshold depends on the scientific purpose that the user has in mind. For example, one could be interested in gathering as many counterparts to eRASS1 as possible, thus maximising the completeness. Others could be interested in creating a smaller but purer sample, thus reducing the number of chance associations as much as possible.  Figure \ref{fig:Meanpurcomp} shows in orange the completeness  for LS10 (right panel), GDR3 (central panel) and CW2020 (left panel), respectively, depending on the \texttt{DET\_LIKE\_0} values of the X-ray sources averaged over the entire eRASS1 survey. Similarly, in blue, the purity curves are constructed by subtracting from 1 the completeness in the random catalogue. This gives the probability that, at a given \texttt{p\_any}, the association is due to chance. 
The purity increases with \texttt{p\_any} (given that it is rare that, among the randoms, a high \texttt{p\_any} is assigned), at the cost of reducing the sample size (completeness). It was already mentioned in \citetalias{Salvato22} for eROSITA/eFEDS that the lower \texttt{p\_any} and lower purity are for sources with low DET\_LIKE\_0, indicating that the X-ray source itself could be a spurious detection. The same trend is found for eRASS1.

The intersection between the completeness and purity curves provides the value of \texttt{p\_any} for which the two quantities are the same.
This is the threshold more commonly adopted \citep[e.g.,][]{Brusa2007, Civano12, Marchesi16} for creating samples of reliable associations.
While this threshold was useful for pencil-beam surveys, where the quality of the data and the number of sources in the primary and secondary catalogues remain the same across the survey area, it is misleading in the case of an all-sky survey, as Figure  \ref{fig:eRASS1density} and \ref{fig:densityfield} have already shown. The consequence is that, depending on the location in the sky, the same \texttt{p\_any} value would refer to different levels of completeness and purity.
In the catalogues that we are releasing with this paper, we provide the \texttt{p\_any} value at the intersection (\texttt{threshold}[6,7,8]) together with the value of the intersection itself (dubbed \texttt{ComPur}[6,7,8]), as a function of \texttt{DET\_LIKE\_0}$>${6,7,8}.
In addition, we provide the value of completeness (\texttt{completeness}[6,7,8]) and purity (\texttt{purity}[6,7,8]) at the \texttt{p\_any} of each source, computed within the eROSITA tile to which the source belongs. An example of how this information can be used for creating a sample of AGN with high purity is described in Section \ref{sec:eRASS1_AGN}.
The maps of \texttt{ComPur}[6,7,8] and corresponding \texttt{threshold}[6,7,8] per each ancillary survey and as a function of \texttt{DET\_LIKE\_0} are reported in Figures \ref{fig:compurMap} for clarity.

Figure \ref{fig:Meanpurcomp} also shows that at the intersection, completeness and purity are higher for LS10, and this happens already at very low values of \texttt{p\_any}, confirming the power of the LS10 model used for determining the correct counterpart. CW2020 reaches high values of purity and completeness, but at a higher value of \texttt{p\_any}, thus substantially limiting the size of the final sample.

\subsection{Separation and magnitude distributions}
\label{sec:Rayleigh}
\begin{figure}
    \centering
    \includegraphics[width=0.4\textwidth]{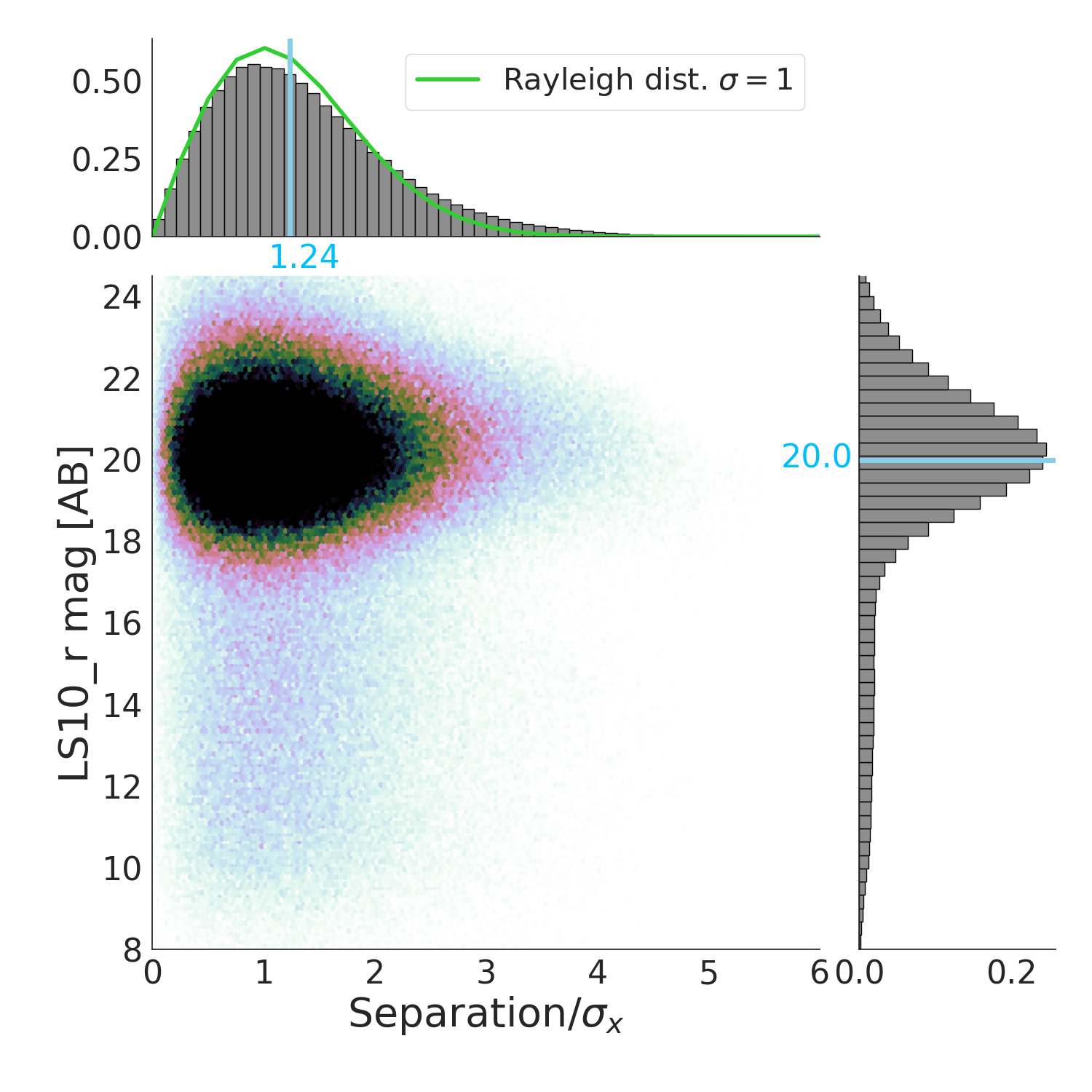}
\includegraphics[width=0.4\textwidth]{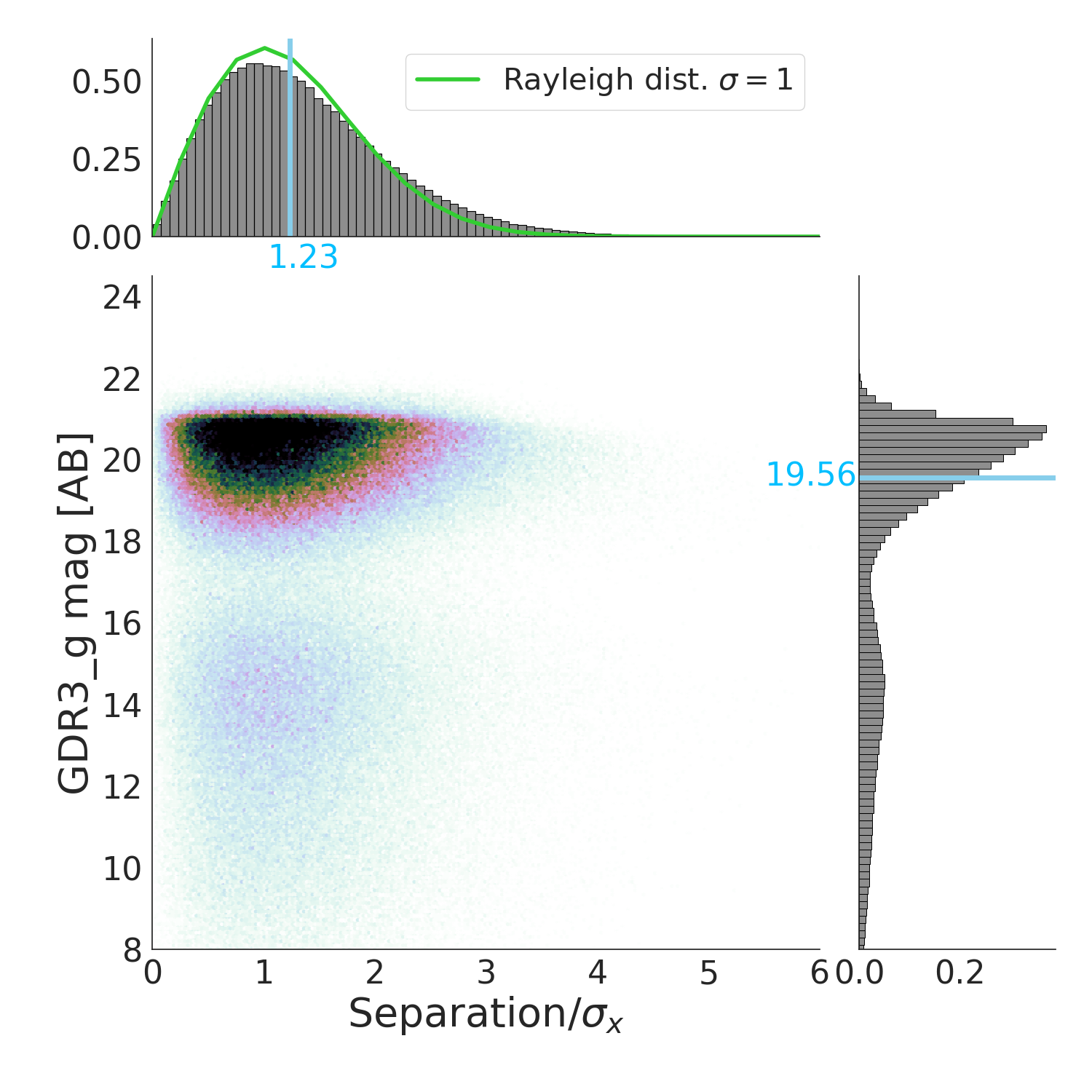}
\includegraphics[width=0.4\textwidth]{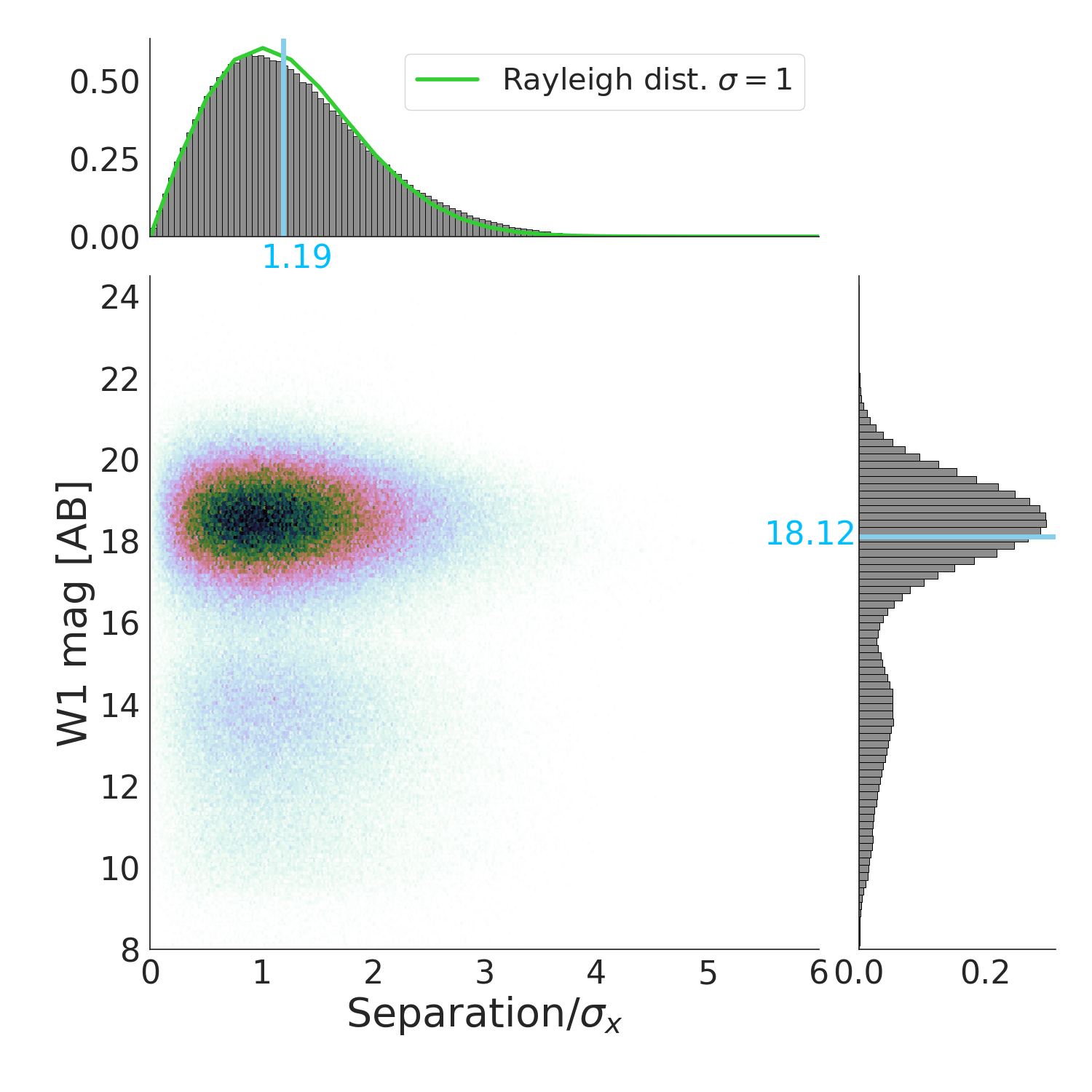}
    \caption{Normalized separation between eRASS1 X-ray position and the selected counterpart as a function of {\it r}-band (LS10), {\it G}-band (GDR3) and W1 (CW2020) magnitude, for sources with secure counterparts (\texttt{p\_any}$>=$threshold6). The separation is normalised by the one-dimensional positional uncertainty of the eRASS1 source. The expected  1$\sigma$ Rayleigh distribution for normalised separations \citep{Rosen2016} is shown in green. Lines indicate the median values of the parameters.}
    \label{fig:Rayleigh}
\end{figure}

The distribution of the observed X-ray to optical/IR separations normalised by the X-ray positional uncertainty is shown in Figure \ref{fig:Rayleigh} as a function of the {\it r}, {\it G}, and W1 magnitudes of the counterparts. The separation between the X-ray position and the assigned LS10, GDR3, and CW2020 counterparts is smaller than 15\arcsec\ in 95\%, 73\% and 95\% of the cases, with a mean of 1.24$\sigma$, 1.23$\sigma$ and 1.19$\sigma$ for LS10, GDR3, and CW2020, respectively.  The faint clouds in all the plots  are corresponding to the brightest objects, most of which are stars. As for eFEDS \citepalias{Salvato22}, there is a trend for larger average X-ray-optical separations at smaller values of \texttt{DET\_LIKE\_0}, because at lower detection likelihood sources typically have larger X-ray positional uncertainty \citep{Brunner2022,  Merloni24}. The distribution is well comparable to the expectation of a Rayleigh distribution with a scale factor equal to 1, shown in green in the plots. This agreement adds confidence to the overall methodology adopted in assigning the counterparts. For GDR3 counterparts, the strongly declining Gaia completeness for objects fainter than G=21 mag imprints a sharp cutoff onto the recovered sample.

\subsection{Limitations of counterpart associations}
\label{sec:limitations}

One point to keep in mind is that the reliability of the counterparts is also bound to the reliability of the ancillary catalogues, which changes depending on the quality/depth of the images and the location in the sky. Artefacts and the presence of very nearby and bright sources will affect the source detection algorithms and, thus, the identification of the counterpart. 
We recommend that the user pay attention to the flags activated on the ancillary catalogues when deciding whether to trust a specific association.

\subsubsection{LS10 quality masks}
While GDR3 and CW2020 are all-sky surveys, LS10 covers only the extragalactic sky. At the border of the survey, the data are incomplete and shallow. The bottom-left panel of Figure \ref{fig:gapa} shows clearly how in these regions the reliability of the association is much lower (high \texttt{p\_any} for the randoms).  Taking advantage of the fact that LS10 provides depth estimates for PSF-like sources for each 0.25 × 0.25 deg2 ‘brick’ and in each filter, we have added two flags:
\begin{itemize}
\item \texttt{InAllLS10}: when the counterpart is a brick where the {\it g, r, i, z} nominal depth (23.5, 23.3, 22.8 or 22.3) is reached in all bands.
\item \texttt{InAnyLS10}: when the counterpart is in a brick where at least one of the {\it g, r, i, z} reached the nominal depth.
\end{itemize}

The masks are indicated in dark-green and magenta colours in the left panel of Figure \ref{fig:densityfield}.

Unlike typical pencil-beam surveys, an all-sky survey naturally includes very nearby, large, and resolved galaxies for which the standard way of measuring optical photometry is not reliable.
For this reason, in our catalogues we have added an additional flag, \texttt{inHEC}, indicating whether or not the eROSITA coordinates fall within the area covered by the galaxy included in the Heraklion Extragalactic CATaloguE \citep[HECATE;\footnote{\url{https://hecate.ia.forth.gr/}}][]{Kovlakas2021}. The catalogue lists about 205k nearby galaxies within a distance of about 200 Mpc, and it is based on the HyperLEDA Database\footnote{\url{http://atlas.obs-hp.fr/hyperleda/}}. Dedicated studies on eROSITA detected sources within galaxies in the HECATE catalogue are presented in \citet{Kyritsis2025} for the relation between X-ray and star formation.

The three columns  (\texttt{inAnyLS10}, \texttt{inAllLS10} and \texttt{inHEC}) are reported in all three catalogues of counterparts to facilitate the comparison.

\subsubsection{eRASS1 sources with zero or multiple counterparts}
\label{subsub:noCTP}
Not all the eRASS1 sources have a counterpart (\texttt{match\_flag}= Null) in each of the three ancillary catalogues, and, at the same time, there are eRASS1 sources for which more than one plausible counterpart is identified within the same ancillary catalogue\footnote{\texttt{match\_flag}= 2 if $\texttt{p\_i}_j/ \texttt{p\_i}_{best}>0.5$, with $\texttt{p\_i}_j$ being the \texttt{p\_i} of the j source in the supporting catalogue}. In our catalogues, all these cases will be listed. Here, it is interesting to look at the reasons for these behaviours.

In the entire eRASS1 catalogue, after removing the flagged sources, there are 10,906 (1.2\%) sources with zero Gaia counterparts within 60\,\arcsec. The LS10 images at the location of these eRASS1 sources typically show the presence of very faint and/or extended sources, which will be missed from the Gaia catalogue, which is shallow and lists only bright compact or point sources.

On the other hand, only 2,267 eRASS1 sources are without CW2020 counterparts (0.2\%). In these cases, the WISE images typically show artefacts (e.g., stripes, saturation spikes) in the WISE/NEOWISE images; these artefacts would prevent the detection of the CW2020 sources, and they would then be missed from the catalogue.

To quantify the number of sources missing LS10 counterparts is complicated by the fact that LS10 does not cover eRASS1 completely. For that, we apply the same \texttt{InAllLS10} and \texttt{InAnyLS10} masks as described in Section \ref{sec:limitations}. All the eRASS1 sources that fall in the footprint of LS10 do have at least one possible counterpart. This is ascribed to the depth and resolution of the data, but also to the strength of the multiwavelength  prior, which is defined using a larger number of features, when compared with GDR3 and CW2020 (see Table \ref{tab:features}). 
In summary, the lack of a counterpart is a direct consequence of the quality of the supporting data.

However, there are also 5,636 sources that have between two and seventeen alternative GDR3 counterparts, with the large majority of the cases located in the Galactic Plane, on the Magellanic Clouds, or in star-forming regions.
Similarly, there are 4,644 sources with a number of possible counterparts between two (65,\% of the cases) and nine counterparts in CW2020. The eRASS1 sources with multiple CW2020 counterparts are once again more concentrated on the Galactic Plane, on the Magellanic Clouds, or in star-forming regions, where the data are confusion-limited, also due to the coarse angular resolution of the CW2020 images. The same is happening for LS10, where 8,933 sources have between two and seventeen counterparts. In all three cases, more than 50\% of the sources have a very low \texttt{p\_any} and \texttt{DET\_LIKE\_0<7}, suggesting the possibility that these X-ray sources are actually spurious detections (see discussion in subsection \ref{subsub:compur}).

\subsection{Comparison between  LS10, GDR3 and CW2020 associations}
\label{sec:Comparison}

While we consider the counterparts determined with LS10 to be more reliable, the data are not available over the whole sky, and it is important to quantify under which assumptions the associations made with GDR3 and CW2020 can be trusted in that region.
For this test, we consider only sources not flagged in the original X-ray catalogue \citep[see details in][]{Merloni24} and outside the area covered by each HECATE galaxy. We cross-matched the LS10 counterparts with those from the GDR3 and the CW2020 catalogues via eRASS ID, and examined how the agreement varies as a function of X-ray detection likelihood, \texttt{DET\_LIKE\_0}, within the \texttt{inAllLS10} and \texttt{inAnyLS10}  footprints. The results are summarised in Table \ref{tab:DR3_CW2020_inAnyLS10} and described in greater detail below. 

\subsubsection{Within the footprint of LS10}
\noindent
At increasing eROSITA detection likelihood, the overall number of sources within the LS10 footprint decreases, but the fraction of sources with the same counterparts in GDR3 and CW2020 increases. This is because there is a correlation between the detection likelihood and the brightness of the X-ray source, which overall corresponds to a smaller positional uncertainty and brighter counterparts. Then, due to the depth of the ancillary data, the agreement is higher between LS10 and CW2020 than between LS10 and GDR3. In this latter case, if we consider only LS10 sources brighter than 20.7 magnitude in {\it G} band\footnote{The magnitude limit of Gaia}, the agreement between LS10 and GDR3 increases up to 94.4\% at \texttt{DET\_LIKE\_0}$>$8. A similar trend is seen for the comparison between LS10 and CW2020, limiting the LS10 sample to sources brighter than 20.4 in W1 (thus simulating the average depth of CW2020). It is reassuring to see the very good agreement between the associations, despite the different features used to identify X-ray emitters in each survey.
Finally, it is  also reassuring that the fraction of sources in agreement is consistent between the area within \texttt{inAllLS10} and \texttt{inAnyLS10}, respectively. This  allows us to conclude that the prior adopted for the determination of the counterpart in the LS10 region is also robust when the quality of the data is not optimal.

\begin{table}
\centering 
\small
\caption{Fraction of sources in GDR3 and CW2020 sharing the same counterpart as identified using LS10 for the \texttt{inAllLS10} region (top) and the \texttt{inAnyLS10} region (bottom).} 

\begin{tabular}{c|c|c|c}
\hline
\hline
& & &\\
\texttt{DET\_LIKE\_0} & \texttt{inAllLS10} & same  & same   \\
&  &  LS10-GDR3  &  LS10-CW2020 \\
 & & &\\
   \hline
   & & &\\

   $\ge6$ & 598,893 & 56.3\,\% & 82.5\,\%  \\
   $\ge6$ \& {\it g}<20.7 &304,566& 93.4\,\% & -- \\
   $\ge6$ \& W1<20.4 &511,222& -- & 89.6\,\%\\
   \hline
   $\ge7$ & 470,757 & 62.4\,\%& 86.3\,\% \\
   $\ge7$ \& {\it g}<20.7 &266,102&94.3\,\% & -- \\
   $\ge7$ \& W1<20.4 &417,013&--&91.6\,\% \\
   \hline
   $\ge8$ & 388,292 & 67.2\,\% &88.8\,\%  \\
   $\ge8$ \& {\it g}<20.7 &237,317& 94.9\,\%& -- \\
   $\ge8$ \& W1<20.4 &352,289& --  & 93.0\,\%\\
\hline
\hline
& & &\\
\texttt{DET\_LIKE\_0} & \texttt{inAnyLS10} & same  & same   \\
&  &  LS10-GDR3  &  LS10-CW2020 \\
   & & &\\
   \hline
   & & &\\

   $\ge6$ &651,935 &55.9\,\%&81.9\,\% \\
   $\ge7$ & 511,742 &62.6\,\%&85.7\,\%    \\
   $\ge8$ & 421,784 &67.3\,\%& 88.2\,\%\\
\hline
\hline
\end{tabular}
\tablefoot{Only the first best counterpart(\texttt{match\_flag}=1) is considered in this comparison, and no cut on \texttt{p\_any} is applied.}
\label{tab:DR3_CW2020_inAnyLS10}  

\label{tab:DR3_CW2020_inAllLS10}      
\end{table}

\subsubsection{Outside the footprint of LS10}
\noindent
For the area outside LS10 (defined by \texttt{inAnyLS10}=0),  the comparison between GDR3 and CW2020 is presented in Table \ref{tab:DR3_CW2020}. 
At any detection likelihood, there are more eRASS1 sources with a counterpart only from CW2020 than from GDR3. This is consistent with the lower reliability of the Random Forrest prior based on GDR3, presented in Section  \ref{section:counterparts}. It means that in many cases, the GDR3 counterparts are only chance associations, and a higher threshold for the purity must be applied.

\begin{table}
\centering 
\small
\caption{Comparison of counterpart associations obtained using GDR3 and CW2020  in the region outside the LS10 footprint.} 

\begin{tabular}{c|c|c|c}
\hline
& & \\
   \texttt{DET\_LIKE\_0} & only CW2020 &  only GDR3 &same CTP\\
   & & \\
   \hline
   \hline
   & & \\
$\ge6$ & 4,401/220,132 & 827/220,132 &58.1\,\%\\
$\ge7$ &3,034/170,244 & 647/170,244 & 63.6\,\%\\
$\ge8$ &2,038/137,566&541/137,566&68.6\,\%\\
\end{tabular}
\tablefoot{We report the number of sources that have a counterpart only from one supporting survey and the fraction of sources for which the counterpart is the same.}
\label{tab:DR3_CW2020}      
\end{table}
\vspace{0.5cm}
\subsection{Comparison with {\sc HamStar}}

\citet{Freund2024} used the Bayesian algorithm {\sc HamStar} \citep{Freund18},  to identifying the GDR3 counterparts to eRASS1 associated  with a coronally emitting star. They found 149,290 reliable associations ($p^i_\mathrm{stellar}$>0.53), out of which 68,614 (58,573) are in the region \texttt{InAnyLs10}(\texttt{inAllLS10}). Of these, 91.8\% (92.9\%) are the same counterparts identified here and classified as Galactic in Section \ref{STAREX}. For the remaining 8.2\% (7.1\%) of the sources, the counterparts provided in this work have high \texttt{p\_i}. In addition, they are  classified as extragalactic, an option that by construction is not considered by {\sc HamStar}. For these sources, it is likely that the X-ray emission is produced by both counterparts. In our catalogue, we will provide a flag indicating whether or not the source coincides with the counterpart determined in \citet{Freund2024}.

\section {Consistency check with ROSAT} 
\label{Rosat}
In \citet{Salvato18a}, the counterparts to the 135,118  ROSAT X-ray sources detected over the whole sky in the newly re-analysed data \citep[ROSAT/2RXS][]{Boller16} were identified using \texttt{NWAY} and a prior based on AllWISE \citep{Wright2010}. The goal here is to quantify the fraction of sources with the same  counterpart as in our eRASS1 catalogues. This is relevant for time domain studies, for which the ROSAT data point can help in identifying long-term variability.
For the match, we have considered the 2RXS-AllWISE counterparts and the eRASS1-LS10, GDR3, and CW2020 counterparts computed here, allowing for a maximum distance of 2\arcsec. We found a match for 22,716, 25,416, and 25,364 sources, respectively. Using  the classification criteria described in the next section, 67\% of the sources in common  with LS10 are extragalactic, while 54\% of the GDR3 sources have a probability higher than 80\% \footnote{using PGAL>0.8 or PQSO>0.8)} to be extragalactic. A star/galaxy classification is not available for CW2020, but using a threshold in W1 and W2 photometry as discussed in Section \ref{subsec:DR1CW2020GP}, 66\% can be considered extragalactic. Interestingly, in all three catalogues, at least 80\% of the matched sources have \texttt{p\_any}$>$0.8, making the associations highly secure. Future works will have the counterparts of ROSAT/2RXS directly recomputed with the method presented in this paper.
At this stage, in the catalogues that we release, two columns with the ROSAT ID from \citet{Boller16} and AllWISE ID \citep[from][]{Wright2010} will indicate whether the source is also associated with the ROSAT/2RXS catalogue.

\section {Galactic and extragalactic Classification}
\label{section:CTPproperties}
In this section, we look at the multi-wavelength properties of the counterparts, intending to classify them as Galactic or extragalactic.

\subsection{Classification in GDR3}
For the catalogues of counterparts determined via Gaia, the user can adopt the Galactic/extragalactic classification from Gaia \citep[\texttt{PQSO}, \texttt{PGAL}, \texttt{Pstar}, \texttt{PWD}, \texttt{Pbin} for the probability to be a QSO, a galaxy, a star, a white dwarf or a binary, respectively][]{GaiaDR3}. 
These probabilities  are included in the catalogue released with this paper for the convenience of the user. When available, the Gaia redshift (either original or from Quaia,\footnote{Sources thought to be AGN based on their Wise colours got their redshift recomputed using a training sample of similar sources with spectra from SDSS. See more in Section \ref{subsec:comparison_quaia_allwise}.} \citealt[][]{Storey-Fisher24}) is also reported.

\subsection{Classification in CW2020}
Unlike GDR3, CW2020 does not provide a star/galaxy classification. However, stars should have mid-IR colours close to zero in the Vega system \citep{Stern2012}. In Section \ref{subsec:DR1CW2020GP}, we further defined the cut and demonstrated that a cut at {\it W1}-{\it W2}$<$0.3 in the Vega system includes most of the Galactic sources with a minimal contamination from extragalactic sources. Nevertheless, there are extragalactic sources with {\it W1}-{\it W2}$<$0.3  as there are Galactic sources with {\it W1}-{\it W2}$<$0.3 (see more details in Section \ref{subsec:DR1CW2020GP} and Figure therein. The users will be able to further clean the selection by using more stringent cuts, or by combining information from other catalogues.

\subsection{Classification in LS10}
For the counterparts determined using LS10, the Galactic/extragalactic nature of the source can be extrapolated using various criteria. \citetalias{Salvato22} demonstrated that, at the X-ray depth of the eROSITA/eFEDS sample, the Galactic/extragalactic nature of X-ray sources can be predicted well from their optical/IR colours; they used a training sample of nearly 8000 spectroscopically classified X-ray sources to define an optimal boundary in {\it g-r} vs. {\it z-W1} colour-colour space that most reliably separates extragalactic and Galactic sources. In particular, the cut creates a complete sample of extragalactic sources, although it includes many stars. Complementarily, the WISE vs X-ray fluxes used in \citet{Salvato18a} were reliable in selecting stars at the cost of missing nearby extragalactic sources. For this reason, we introduce here STAREX, a more efficient machine learning algorithm that we developed to distinguish Galactic sources from the extragalactic ones, using Legacy Survey DR10 data.

\begin{figure*}
    \centering
\includegraphics[width=12cm]{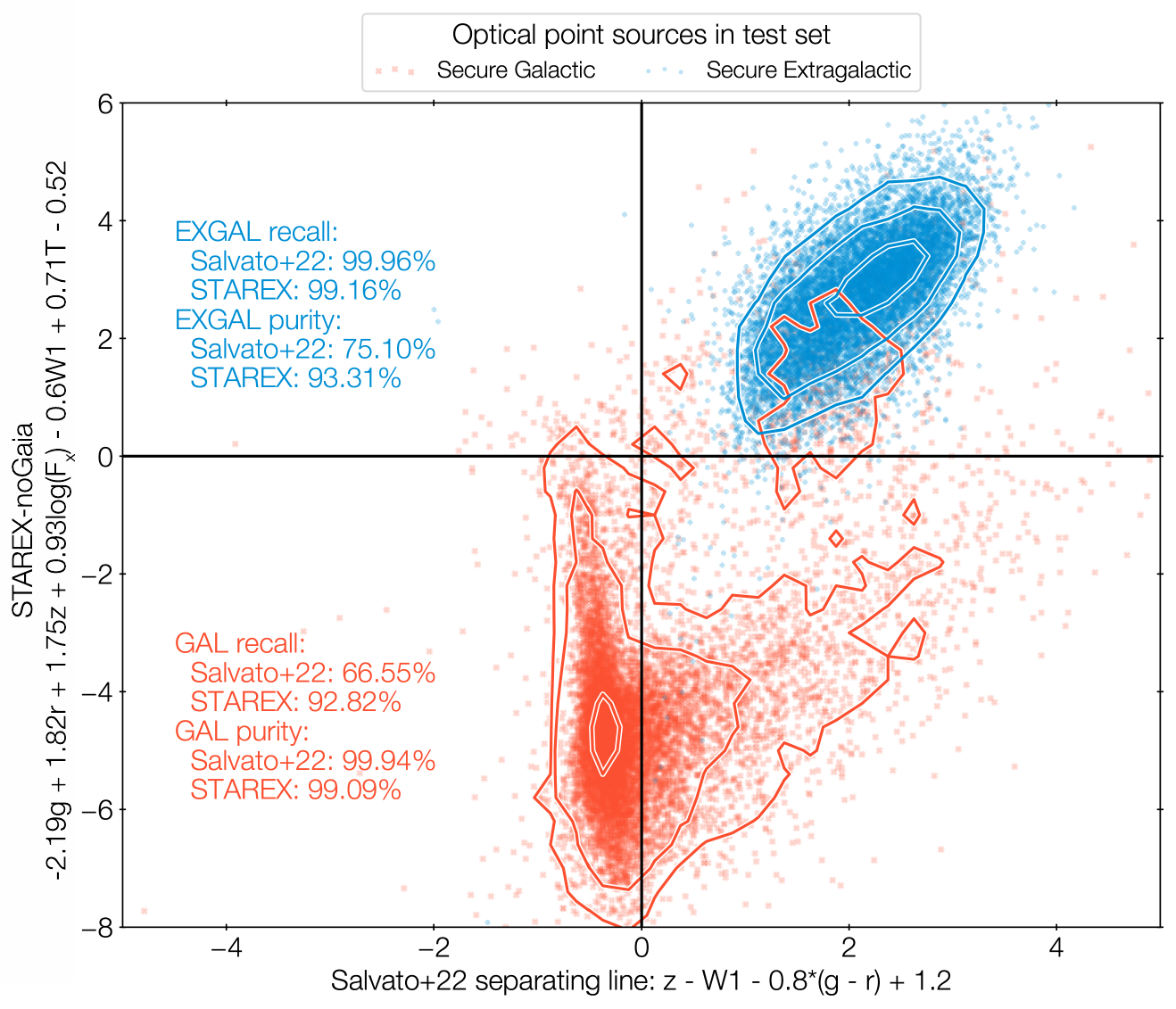}
    \caption{Classification performance on the test sample, restricted to sources appearing point-like in LS10 images. The horizontal axis shows the distance from the \citetalias{Salvato22} photometric classification criterion (with >0 indicating extragalactic and <0 Galactic classification). The vertical axis shows our new separating line, for the case where no Gaia information is available (STAREX-noGaia), using the same magnitudes but also the optical type (0 here for point-like) and the soft-band X-ray flux.
    The recall for secure extragalactic (blue) and Galactic (red) test sources is reported in the left panels, followed by the corresponding purity values.
     The contours enclose 25\,\%, 75\,\% and 90\,\% of the sources. 
    }
    \label{fig:classification_purity}
\end{figure*}

\begin{figure}
\includegraphics[width=\columnwidth]{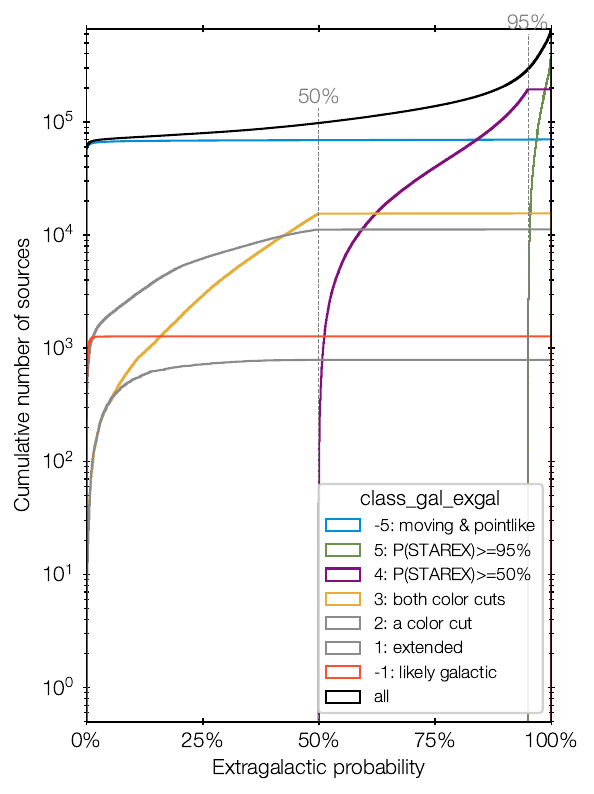}
    \caption {Cumulative distribution of sources within the LS10 footprint  sorted in order of reliability of the star/galaxy classification. Sources with a STAREX probability to be extragalactic below 50\,\% are further split into classes (from 3 to -1) using a combination of STAREX output and diagnostic colour-colour diagrams (grzW1 and W1X) as described in \ref{subsec:LS10_class}.}
    \label{fig:LS10_class}
\end{figure}

\subsubsection{STAREX}
\label{STAREX}
We have created a training sample of 131,000 eRASS1 sources, split almost 50/50 between Galactic and extragalactic objects. The former are defined as secure eRASS1 counterparts with PSTAR >90\%  in GDR3, while the extragalactic sources have a secure spectroscopic redshift z>0.002 from DESI \citep{DESI2025}. For this (almost) ground truth X-ray sample, we have retrieved the LS10 catalogue information.

The optimal separation between Galactic and extragalactic sources is found with machine learning. We tested Fisher’s discriminant analysis,  random forest and logistic regression. The features used for classification are an expanded set of those used in \cite{Salvato22}. This includes the AB magnitude from LS10 in the {\it g,r,z,} and W1 bands, and the morphological type which is encoded as integers from 0 to 5 (\texttt{PSF, REX, EXP, DEV, COMP, SER}) by the complexity of the best-fit profiles \cite[see][]{Dey2019}. We also included the log of X-ray flux (in erg/s/cm$^2$ in the 0.2-2.3\,keV band as available from eROSITA). For sources in the training sample detected by Gaia, parallax, inverse variance, mean magnitude in BP, RP, and G as listed in LS10 are also included. These features are then used to train a supervised machine-learning classifier. However, not all LS10 sources have Gaia information due to different depths. To take advantage of this information when available, we train two classifiers, one appropriate for sources with Gaia information (11 input features) and another that can be applied to any X-ray LS10 source (6 input features). Their training is otherwise identical. 

Training proceeded as follows: The ground truth sample was split into a training set (75\% ) and a test set (25\% ), the latter used for validation. The training-validation data is normalised to have a mean of zero and a standard deviation of one, and the normalisation information is stored for application to the test data and later application. The training-validation data is then split with K$=$10 folds into training and validation data sets
For each of these ten training samples, a classifier is then trained on the input features and ground truth label (galactic=0, extragalactic=1). For simplicity and interpretability, we chose logistic regression, where the probability of the extragalactic class given input feature vector $x$ is:
\[
p(x)=\frac{1}{1 - e^{\alpha + \beta x}}=\frac{1}{1 - e^{- (x - \mu)/s}}
\]
The weight vector $\mu$ quantifies the importance of each feature, and the scale parameter $s$ determines the sharpness of the probabilistic transition from one class to the other. Equivalently, the linear dependence can be quantified with intercept $\alpha=\mu/s$ and weights $\beta=1/s$.
Training then maximises the log-likelihood function:
\begin{equation}
\log{\cal{L}}(\alpha,\beta) = w_\mathrm{gal  }\sum_{i \in \mathrm{gal}}\log(1-p(x_i))
              + w_\mathrm{exgal}\sum_{i \in \mathrm{exgal}}\log p(x_i)
\end{equation}
The best-fit parameters (intercept $\alpha$, weights $\beta$) of the ten classifiers for each fold are then combined by averaging. This gives a final logistic classifier less dependent on the specific training sample. 
The linear separation planes ($0= \alpha+\beta x$) defined with and without the Gaia features, are:\\

\begin{multline}
0 = 0.14\,{\rm TYPE} - 2.25g + 0.96r - 1.60{\rm w1} + 1.58z - 4.54bp + \\2.16G+ 4.07rp+ 1.17\log_{10}(F_{0.5-2 keV})- 0.47\mu + 0.0\mu\_ivar + 13.73
\end{multline}
and 
\begin{multline}
0 = 0.71\,{\rm TYPE} - 2.19g + 1.82r - 0.60{\rm w1} + 1.75z + \\
0.93 \log_{10}(F_{0.5-2 keV}) - 0.52
\end{multline}
where TYPE corresponds to the morphological type (0 to 5).
\begin{table}
    \centering
    \caption{Source classification performance on the test set.}
    \begin{tabular}{ll|ll}
        \hline
        Subset & metric & S22 & STAREX-noGaia \\
        \hline
        \hline
        point sources & recall exgal & 100.00\% & 99.10\% \\
                      & purity exgal & 69.00\% & \textbf{95.64\%} \\
                      & recall gal & 67.67\% &\textbf{96.75\%} \\
                      & purity gal & 100.00\% & 99.34\% \\
        \hline
        all sources   & recall exgal & 99.97\% & 99.13\% \\
                      & purity exgal & 71.84\% & \textbf{95.37\%} \\
                      & recall gal & 67.03\% & \textbf{97.34\%} \\
                      & purity gal & 99.96\% & 99.24\% \\
        \hline
        w. Gaia info. & metric & S22 & STAREX-Gaia \\
        \hline
        point sources & recall exgal & 99.96\% & 99.16\% \\
                      & purity exgal & 75.10\% & \textbf{93.31\%} \\
                      & recall gal & 66.55\% & \textbf{92.82\%} \\
                      & purity gal & 99.94\% & 99.09\% \\
        
        \hline
        all sources   & recall exgal & 99.72\% & \textbf{96.60\%} \\
                      & purity exgal & 77.91\% & \textbf{91.39\%} \\
                      & recall gal & 63.31\% & \textbf{93.31\%} \\
                      & purity gal & \textbf{99.44\%} & 95.24\% \\
        \hline
        \hline
    \end{tabular}
    \label{tab:classifier_performance}
\end{table}

Figure~\ref{fig:classification_purity} illustrates the performance of our new separating line for the case without Gaia information (which is the case for most of the eRASS1 sources) on the vertical axis. For comparison, the horizontal axis of Figure~\ref{fig:classification_purity} shows the \citetalias{Salvato22} separating criterion. 
As already shown in the original paper, the distribution of  the Galactic sources, in red, extends into the extragalactic region. These sources are the reason for a modest recall of Galactic sources.
All statistics comparing \citetalias{Salvato22} and STAREX are listed in Table~\ref{tab:classifier_performance}, including the recall (selection completeness) and purity, for optical point-sources and all sources.
The separation with STAREX is better, with high recall and sample purity both above 95\,\%. For extragalactic sources, the recall is essentially complete with both methods (>99\,\%), but STAREX increases the purity from 69\,\% to 95\,\%. The performance of STAREX further improves when Gaia information is available and used\footnote{Note that the samples used on the top and bottom of Table~\ref{tab:classifier_performance} are different}.

\subsubsection{Final Galactic/extragalactic classification for LS10 sources}
\label{subsec:LS10_class}
In addition to STAREX or the line separators introduced \citetalias{Salvato22} and \citet{Salvato18a}, we also use the information on parallax ($\mu$) and proper motion ($\pi$) provided by Gaia alone to  identify secure Galactic sources.
In particular, we consider 
\begin{multline}
SNR\_\pi = \pi * sqrt(\pi\_IVAR)
\end{multline}
\begin{multline}
SNR\_\mu =sqrt(\mu_{\rm RA}^2 \times (\mu_{\rm RA}\_IVAR)+(\mu_{\rm DEC}^2 \times \mu_{\rm DEC}\_IVAR))
\end{multline}
The same two quantities, computed for the 206k sources in the DR16Q catalogue \citep{Lyke2020} detected by Gaia and with visually inspected redshifts, show that 99.95\% of the sources have either \texttt{SNR\_parallax}$>$5 or \texttt{SNR\_PM}$>$5. We use these thresholds combined with \texttt{TYPE}=PSF to select secure Galactic sources based on their motions.

There are cases where STAREX, the line separators from {\it grzW1} and/or W1X, and the proper motion and parallax from Gaia provide conflicting Galactic/extragalactic classification.
 For this reason, we introduced the quantity \texttt{class\_gal\_exgal}, which provides information on the reliability of the classification.
 
 With \texttt{class\_gal\_exgal} = $-5$ we identify sources that are classified as Galactic because they move  and are point-like (\texttt{TYPE}=PSF). Sources with a STAREX probability of being extragalactic above 95\% (50\%) have  \texttt{class\_gal\_exgal}= 5(4). For STAREX probability lower than 50\,\% we assign \texttt{class\_gal\_exgal}=3(2) to the sources classified as extragalactic based on both (at least one) line separators. Finally, we assigned \texttt{class\_gal\_exgal}$=-1$ to the remaining sources. To summarise, negative numbers indicate stars, positive numbers indicate AGN, and higher absolute values indicate higher confidence.

To double-check our final classification, we first looked at the internal consistency. Only 191 sources in  the area within \texttt{inAllLS10} are considered Galactic (\texttt{class\_gal\_exgal}$=-5$), but STAREX would have classified them as extragalactic.  It is known that compact objects, such as Cataclysmic variables (CVs), occupy the same locus of AGN in many parameter spaces, such as those used in our classification. In fact, the large majority of these sources are either already classified or are candidate CVs following \citet{Schwope2024b}, with which we have  a 98.6\%  agreement in the identification of the counterparts. An additional test was done, checking the spectroscopic redshifts from DESI DR1 \citep{DESI2025} when available. There are about 58,000 eRASS1 sources in the footprint of LS10 with a DESI redshift. Among the 86,599 eRASS1 sources classified as Galactic within the footprint of LS10, 1,274 have a reliable redshift from DESI, out of which 79 have redshift z$>$0.002. On the other hand, there are 569,778 eRASS1 sources classified as extragalactic, 57,162 with redshift from DESI, with only 114 of them having z$<$0.002. A visual inspection of these cases reveals, for the most part, sources very close to a saturated star to which the DESI spectrum is probably related.

Figure \ref{fig:LS10_class} shows the breakdown of the best LS10 counterpart to the eRASS1 point sources (after removing those flagged in \citealt{Merloni24}) in the various classes, as a function of the probability to be extragalactic from STAREX. To further subclassify the sample, we add further information. Firstly, we add the column \texttt{class\_jetted}, which indicates whether the LS10 position is in proximity to a blazar (jetted=1) in the BzCAT \citep{Massaro2015}, a flat-spectrum radio source (jetted=2) in CRATES \citep{Healey2007}, or a blazar in Simbad (jetted=3), or 0 otherwise.
We added the column \texttt{simbad\_known\_galactic} to report whether a Simbad positional match within 1\arcsec\ identified a source with main type X-ray binary, cataclysmic variable, High-mass or Low-mass X-ray binary.

\section{Redshifts}
\label{sec:redshifts}
For all the eRASS1 sources classified as extragalactic in the entire LS10 footprint, we also provide redshifts or redshift estimates.
First, we used the LS10 coordinates to perform a match (maximum distance of 1\arcsec) to a large compilation of spectroscopic redshifts taken from the literature  \citep[Igo et al., in prep]{Kluge2024}, including, among others: Quaia \citep{Storey-Fisher24}, 2dF QSO catalogue \citep{Croom09}, 6dFGS \citep{Jones09} and the latest release of DESI DR1 \citep{DESI2025}. 
The compilation has many astrophysical objects with more than one redshift reported in the literature, with redshift values in disagreement. This is often due to the inclusion of lower-quality spectra or due to differences in the way redshifts have been derived from the spectra. We decided to remove all those sources in the compilation with redshifts in disagreement larger than 0.1. This is done for two reasons. First, we are not in the position to decide which redshift is the correct one. Secondly, we use the spectroscopic sample to create a clean reference sample to measure the fraction of outliers and the accuracy for the photometric redshifts that we computed with the algorithm {\sc{Circlez}} \citep{Saxena2024} for the sources without spectroscopic redshift. In addition to the cleaned public compilation, we added new 137 redshifts acquired by eROSITA members over the years with various instruments at different telescopes and 217 redshifts that are provided by eROSITA collaborators from ongoing ESO campaigns (Scialpi et al., in prep).
In total 196,083 LS10 counterparts to eRASS1 have a spectroscopic redshift. For the rest of the sources, we rely on photometric redshifts.

\begin{figure*}
    \centering
    \includegraphics[width=0.45\linewidth]{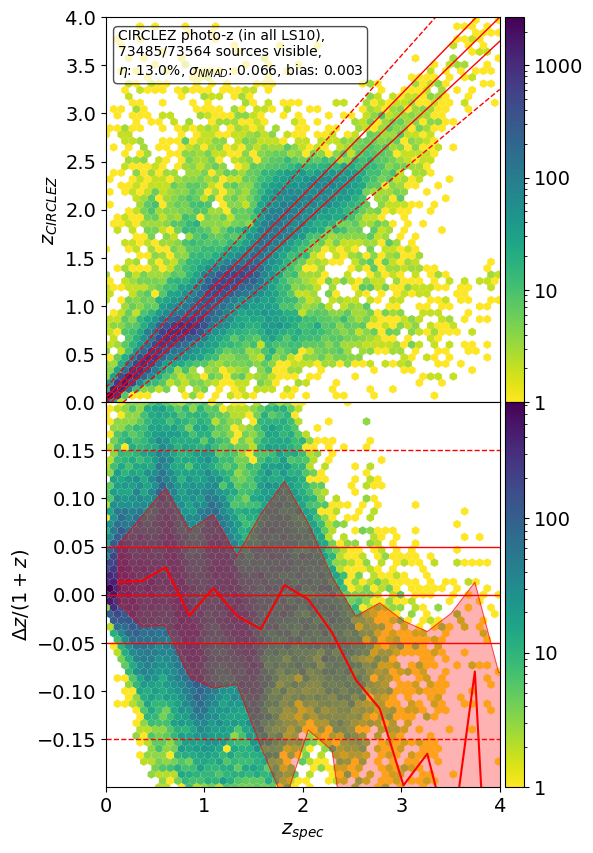}
    \includegraphics[width=0.45\linewidth]{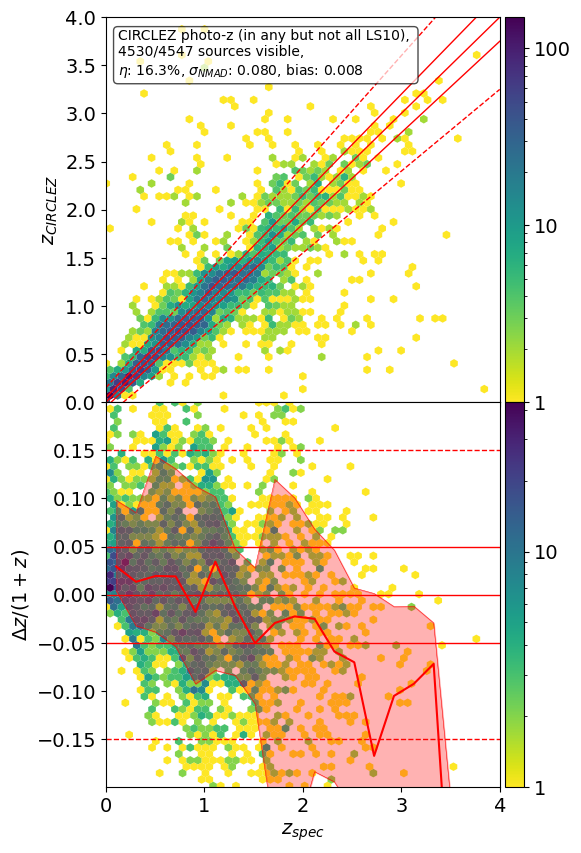}
    \caption{Comparison between photometric and spectroscopic redshifts for  sources within (left panels) and outside (right panels) the inAllLS10 area. In the top panel, the lines represent the one-to-one relation and the thresholds adopted to define outliers (see text and footnotes). The bottom panel shows the normalized residuals $(z_{phot}-z_{spec})/(1+z_{spec})$. The red lines in the lower panels indicate the moving medians and the 1 sigma region, revealing a negative bias above redshift 2. See main text for more details.}
    \label{fig:zphotzspec}
\end{figure*}

Deriving photo-z for AGN is notoriously difficult, due to the a priori unknown host/active nucleus contribution at a given wavelength. 
Classical photometric redshift techniques use the total fluxes of a source and explicitly (SED fitting) or implicitly (machine learning) the colour-redshift relation characteristic of a source to determine the redshift \citep[see][for a review]{Salvato18b}. {\sc{Circlez}} is a machine learning algorithm based on fully connected neural network (FCNN) that estimates the redshift using as features, not only the total fluxes, but also the fluxes and the residuals after subtracting the best fitting morphological type\footnote{Described in \url{https://www.legacysurvey.org/dr10/description/}}, in addition to the aperture photometry and colors within apertures.
Here we used the same model that was defined in \citet{Saxena2024}  using a sample of 14,000 X-ray detected extragalactic sources with secure spectroscopic redshift. A direct comparison between the spectroscopic and photometric redshift is shown in Figure \ref{fig:zphotzspec}. For the 74,089 sources in the \texttt{inAllLS10} with spectroscopic redshifts, we obtained an accuracy\footnote{$\sigma_{NMAD}= 1.48* median(\Delta z/(1+z_{spec}))$} of 0.067, with only 12.7\%  outliers fraction\footnote{Defined as fraction of sources $\eta$ with $|z_{\rm phot} - z_{\rm spec}|/(1+z_{\rm spec})>0.15$}. Worth noticing is the fact that the bulk of the outliers are close to the 0.15 separation line used in literature, but arbitrarily defined.
The accuracy and the fraction of outliers are very similar at any detection likelihood, and it is superior to the one reached in eROSITA/eFEDS \citepalias{Salvato22}, thanks to the new approach implemented in {\sc Circlez}.
In fact, for the sources outside the \texttt{inAllLS10} area (i.e., the region with shallower LS10 data), the accuracy (0.078) and bias (0.001) are very similar to those with the best photometry, and only the fraction of outliers (15.4\,\%) increases slightly. 
In addition, many of the sources at $z_{\rm phot}>3$, are outliers, due to their faintness and larger photometric errors. To interested readers, this is discussed in detail in \citet{Saxena2024} and \citet{Roster2025}. For these type of sources, rather that the photometric redshift, the redshift distribution function (PDZs), should be used.  Unlike PDZs from SED fitting that are too optimistic  (i.e., producing errors that are too narrow also for sources that are known outliers, \citep[e.g.,][]{Brescia2019}), the PDZs obtained with {\sc Circlez} are realistic \citep[see][]{Saxena2024}. The PDZs for eRASS1 are available upon request.
Only for 454 (about 0.008\,\%) of the sources classified as extragalactic in LS10, the photometric redshift could not be constrained due to a lack of sufficient information (i.e., low number of photometric points).

\section{Defining eRASS1 AGN samples}
\label{sec:eRASS1_AGN}
We release with this paper the counterparts to all point sources in the main catalogue of eRASS1, identified using LS10, GDR3, and CW2020, respectively (see next section). The users will then be free to select their subsamples of sources based on any (combination of) criteria (e.g., completeness, purity, depth/quality of X-ray or ancillary data). 

The more basic sample of AGN could be generated by simply selecting all the sources that are classified as extragalactic. However, this would create an inhomogeneous sample of AGN in LS10, given that the depth of the ancillary data varies over the sky. Alternatively, controlled AGN samples can be generated by accounting for the quality of the ancillary data and reliability of the X-ray detection. In this section, we follow the second path and create controlled samples of AGN using the counterparts from LS10, GDR3, and CW2020 to understand the complementarity of X-ray-selected AGN with other selection methods.

As an example, we define 6 high-purity samples following the steps listed below and represented in Figure~\ref{fig:FlowChart}, together with the number of sources remaining after each step. We describe here the construction of samples 1 (purity level higher than 90\,\%, everywhere in LS10), 5 (AGN selected outside the footprint of LS10 using GDR3, with same completeness and purity, for eRASS1 sources with \texttt{DET\_LIKE\_0}>8)   and 6 (same as 5 but using cw2020). Samples 2, 3, and 4 (using LS10, with same completeness and purity, for or eRASS1 sources with \texttt{DET\_LIKE\_0}>{6,7,8])   are described in Appendix \ref{appendix:LS10AGNSample}. We will then compare the Sample 1 of AGN  with those selected using AllWISE, Gaia, and a combination of WISE and Gaia.

\begin{figure*}
\centering
\includegraphics[width=\textwidth]{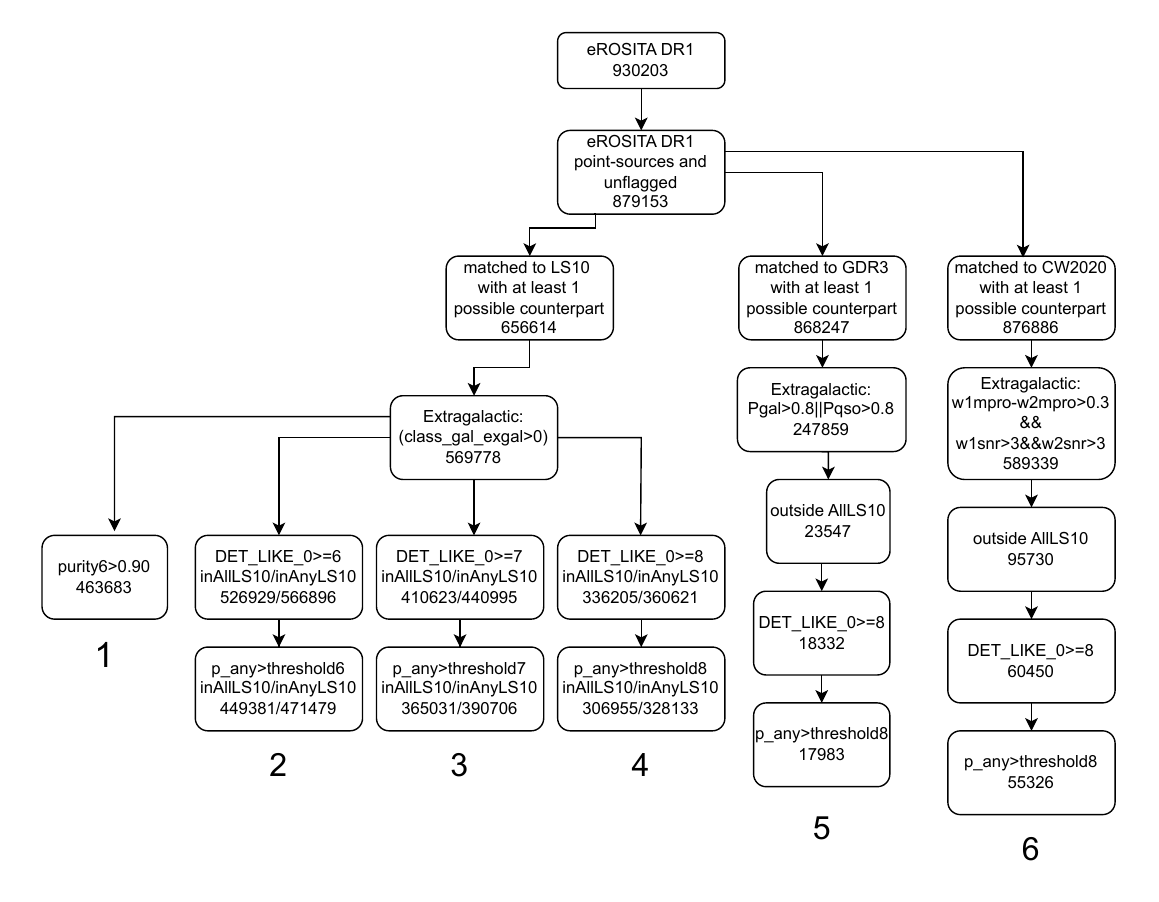}
\caption{Flowchart describing the construction of six samples of candidate AGN with different supporting data, on different regions and with different levels of completeness and purity. For the detailed construction of the samples, see Section~\ref{sec:eRASS1_AGN}.)}
\label{fig:FlowChart}
\end{figure*}

\subsection{eROSITA AGN within LS10}
\label{subsec:DR1LS10}
We start to construct our AGN catalogue from the point-like (\texttt{EXT\_LIKE}=0), unflagged \citep[\texttt{FLAG\_{*}}=0, see][]{Merloni24} in the X-ray catalogue  and with the counterpart from LS10 classified as extragalactic (\texttt{class\_gal\_exgal}$>$0). 
A first selection (Sample 1 in Figure~\ref{fig:FlowChart}) could be applied by taking into account the spatial dependence of the purity of crossmatches (see Section~\ref{sec:limitations}) by applying a local \texttt{p\_any} threshold (set for each eROSITA tile),  such that the purity[6] is greater than $>$90\%}. In such a sample, the population would differ from tile to tile, depending on the depth of both the eRASS1 and the LS10 available there. We call this sample DR1\_LS10\_EXGAL\_DET6P90.

\begin{figure}
\centering
\includegraphics[width=\columnwidth]{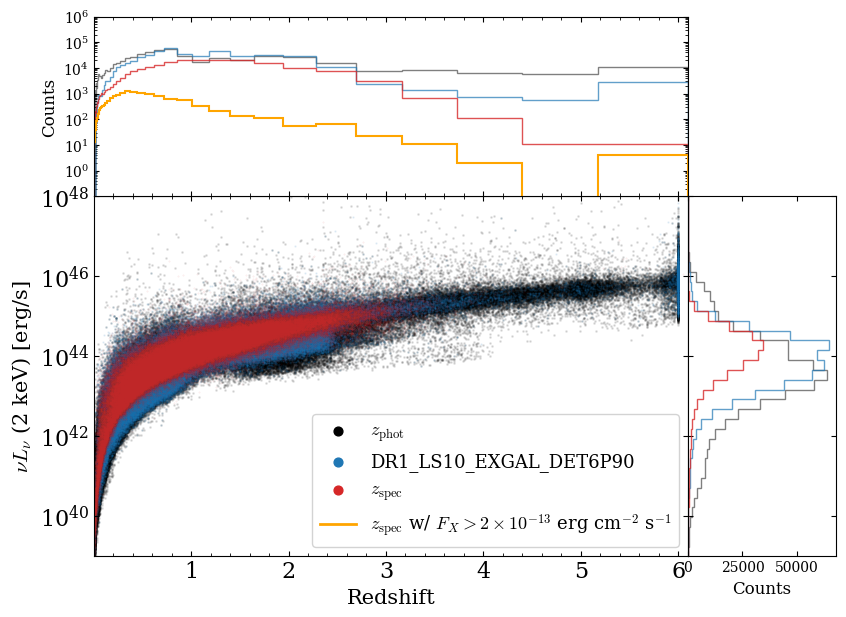}
\caption{X-ray (2keV) luminosity-redshift distribution for eRASS1 sources within the LS10 footprint, compared to the same distribution from the high purity and reliability AGN sample DR1\_LS10\_EXGAL\_DET6P90 (blue). Symbols indicate the redshift:  spectroscopic (red) and photometric (black). The apparent excess of sources at z = 6 is attributed to unreliable photometric redshift estimates (see Section~\ref{sec:redshifts}). For completeness, the histogram distributions for the redshift (top) and the X-ray luminosity (right) are also shown. In the redshift distribution we also show, in orange, the spectroscopic distribution for sources at the depth of ROSAT ($F_{\rm 0.5-2 keV}>2\times10^{-13}$ erg s$^{-1}$ cm$^{-2}$).}
\label{fig:LxZ}
\end{figure}

Figure~\ref{fig:LxZ} shows the X-ray luminosity and redshift distribution for the sample DR1\_LS10\_exgal\_DET6P90, where photometric redshifts are used for the 383,349 sources without spectroscopic redshift (75,496; see Section \ref{sec:redshifts}).
The figure suggests that a limited fraction of sources in the sample are at luminosity lower than $10^{41}$ erg s$^{-1}$, a typical threshold adopted for separating galaxies powered by AGN or star formation \citep[e.g.,][]{Lehmer2012, Brightman2011, Bauer2004}. Only a detailed analysis will determine the dominant contribution for these sources.

\begin{figure}
\centering
\includegraphics[width=0.5
\textwidth]{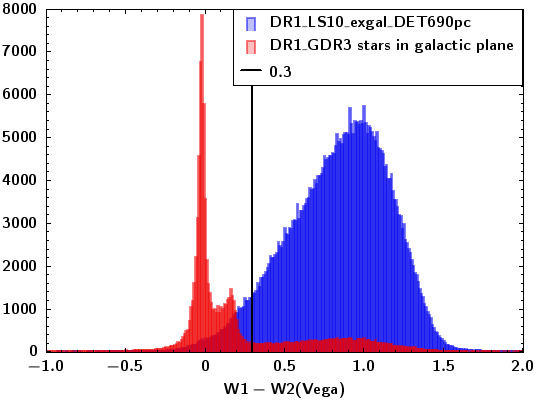}
\caption{CW2020 W1$-$W2 (Vega) colour distribution for the extragalactic sources in the clean extragalactic sample within LS10 (blue) and a clean sample of Galactic sources in the Galactic plane from the GDR3 counterparts to eRASS1. The vertical line indicates the cut to apply to define a sample of extragalactic sources among the CW2020 counterparts to eRASS1, least contaminated by Galactic sources.}
\label{fig:W1W2cut}
\end{figure}

\subsection{eRASS1 AGN in the Galactic Plane using GDR3}
\label{subsec:DR1GDR3GP}
The scheme used for the construction of this sample is represented in Figure~\ref{fig:FlowChart}, following the path for Sample 5.
Because the prior adopted for determining the counterparts using GDR3 is less reliable than for LS10, we decided to consider only sources with \texttt{DET\_LIKE\_0}$\ge$8, so that the positional errors are typically smaller, limiting the area search for the correct counterpart. For the area outside LS10 (\texttt{inAnyLS10}=0), we consider as extragalactic the sources that have a probability of being a galaxy (\texttt{Pgal}) or a QSO (\texttt{Pqso}) larger than 80\% from GDR3. Above the threshold that identifies the \texttt{p\_any} point at which purity and completeness are maximal,  there are 17,983 sources, mostly on the Galactic Plane.
The mean value of purity and completeness of the sample at the \texttt{threshold8} (see Section \ref{section:counterparts}) values is 74.4\%. We will call this sample DR1\_GDR3\_EXGAL\_GP\_DET8CP74.

\subsection{eRASS1 AGN in the Galactic Plane using CW2020}
\label{subsec:DR1CW2020GP}
The scheme used for the construction of this sample is represented in Figure~\ref{fig:FlowChart}, following the path for Sample 6.
The AllWise W1$-$W2\footnote{Vega system} colour cut at 0.7 \citep[e.g.,][]{Assef2013, Assef2018}  is usually adopted for selecting AGN in the Mid-infrared. However, the cut does not fully account for the fact that the instrument has a coarse angular resolution (6.1\,\arcsec in W1 and 6.4\,\arcsec in W2, with an image  pixel scale of 1.375\arcsec) and that the source detection had limited deblending capability. Thus, more than one source could be contributing to the mid-infrared flux. This is even more true at the depth of CW2020 (0.8 and 1.6 magnitude deeper than AllWISE in W1 and W2, respectively), and in the Galactic Plane, where the number of sources reaches the confusion limit.

To define a W1$-$W2 cut based on CW2020 photometry, able to separate candidate AGN among the CW2020 sources associated with eRASS1, we matched the clean sample of extragalactic sources from LS10, DR1\_LS10\_exgal\_DET6P90 to CW2020 with a radius search of 3\,\arcsec, considering only those with a unique counterpart (92\%  of the sample). We repeated the procedure with a clean sample of stars (\texttt{PSS}>0.9) in the Galactic Plane from the DR1\_GDR3 sample. The distribution of the sources in W1$-$W2 is shown in Figure \ref{fig:W1W2cut}. A cut at W1$-$W2>0.3 provides an extragalactic sample that is 95\% complete (only 5\% of the DR1\_LS0\_exgal\_DET6P90 sample have  W1$-$W2<0.3.  However, the sample is not pure as 25\% of the stars in DR1\_GDR3 have  W1$-$W2>0.3.
The CW2020 counterparts to eRASS1 with W1$-$W2$>$0.3, with SNR in W1 and W2 larger than 3, outside the LS10 area, with \texttt{DET\_LIKE\_0}$>$8 and with \texttt{p\_any} above the threshold are 55\,326 and will be considered extragalactic. The mean value of maximum purity and completeness reached at the threshold for \texttt{DET\_LIKE\_0}$>$8 is 82.8\%. We will call this sample DR1\_CW2020\_EXGAL\_GP\_DET8CP83.

\begin{figure}
\centering
\includegraphics[width=0.5\textwidth]{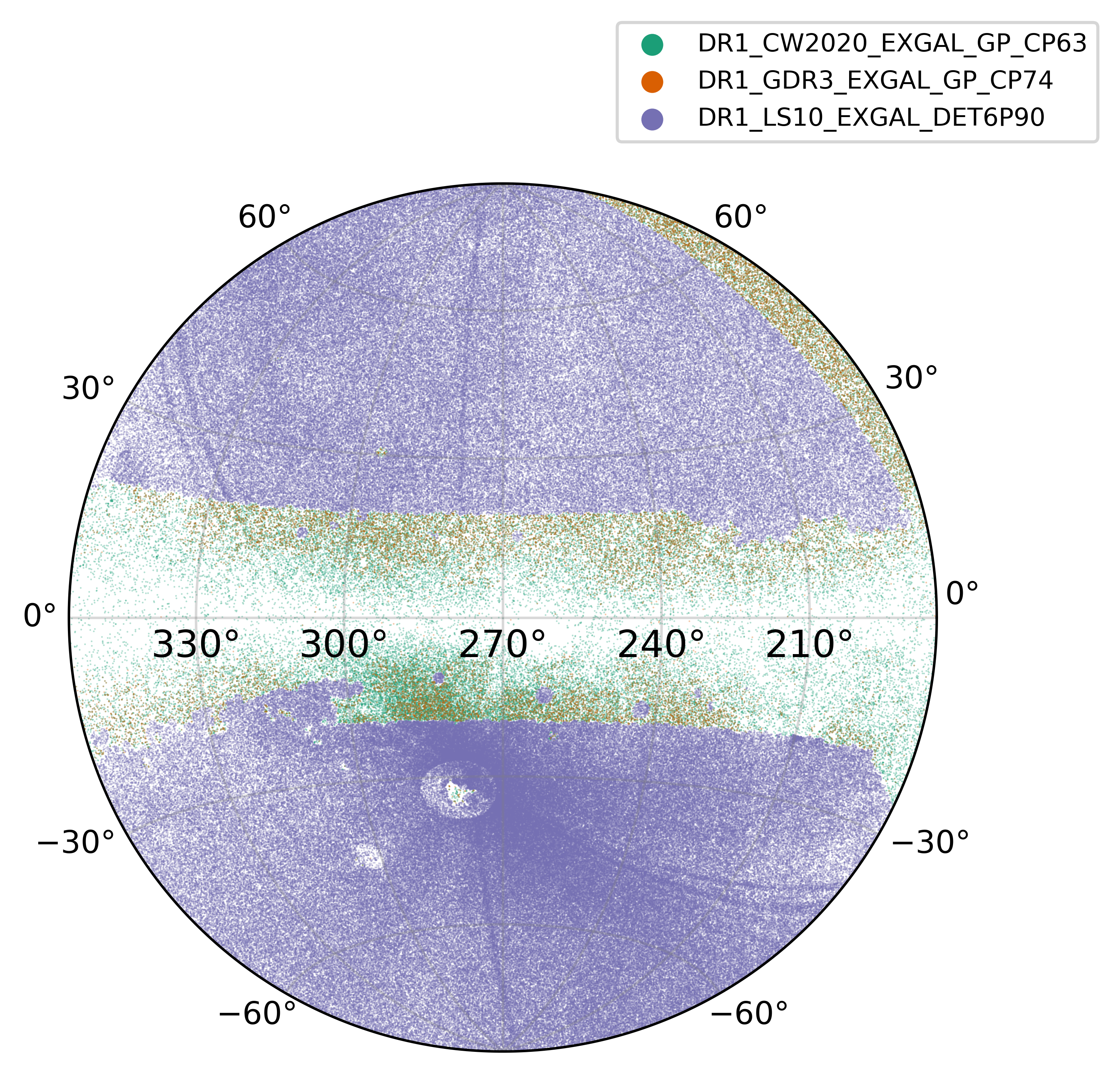}
\caption{Sky distribution of the AGN samples 1 (LS10; purple), 5 (GDR3; orange) and 6 (CW2020; green). Per constructions, the AGN samples from GDR3 and CW2020 occupy the area outside the footprint of LS10. 
}
\label{fig:AGNsky}
\end{figure}

Figure~\ref{fig:AGNsky} shows the distribution of the three AGN samples in the sky: compared to the shallow GDR3, CW2020 does allow great access to low Galactic latitudes, but large gaps in coverage are evident, also due to source confusion.

\subsection{Comparison between eRASS1 AGN within the LS10 footprint and those selected in Gaia, Quaia, and AllWise AGN}\label{subsec:comparison_quaia_allwise}
Now that we have defined clean subsamples of eRASS1 extragalactic sources that are bona fide AGN, selected in the area with the best data quality, we can compare their number density and basic properties with the AGN identified at other wavelengths, such as in the mid-infrared \citep[with AllWISE,][]{Assef2018}, and in the optical \citep[Gaia][]{GaiaDR3} and in mid-infrared and optical simultaneously \citep[with Quaia,][]{Storey-Fisher24}. 

The AllWISE sample of candidates AGN (R90) relies on a cut in W1$-$W2 colour determined empirically using a sample of spectroscopically confirmed, optically bright AGN originally detected in the IRAC channels Ch1 and Ch2 \citep[very similar to W1 and W2 in WISE;][]{Stern2005, Stern2012}. The mid-infrared selection is less biased toward obscured AGN, e.g., those presenting narrow emission lines in the optical. However, at increasing depth of the WISE data, the fraction of star-forming galaxies increases, making the selection less pure. \citet{Assef2018} provides four catalogues with completeness (C) and reliability (R) at 75\% or 90\%. For the purpose of the comparison, we are interested in purity more than in completeness, and for this reason, we selected the R90 catalogue. In the footprint of LS10, the R90 catalogue includes 2,113,681 sources.

\noindent We consider as Gaia AGN sample those listed in the catalogue of Gaia QSOs with a probability to be QSO (\texttt{Pqso}) higher than 80\% (see Section \ref{subsec:Gaia}). This is because the original sample is highly complete but with a purity of only 52\% \citep[][]{Gaia2023b}. The sample includes $\sim$2/3 of the original 6.6M sources, further limited to 1,437,708 when considering only those in the footprints of LS10.

\noindent The Quaia AGN sample is formed by the subsample of Gaia QSO (regardless of the value of \texttt{Pqso}) that is also selected in unWISE. The catalogues include
755,850 quasar candidates with G $<$ 20.0 and 1,295,502 with G$<$20.5.
For these sources, the redshifts are determined on GAIA spectra using machine learning trained on similar sources with spectra from SDSS. 
The Quaia sample in the footprint of LS10 lists 603\,639 sources. \\

\begin{figure}
\centering
\includegraphics[width=0.5\textwidth]{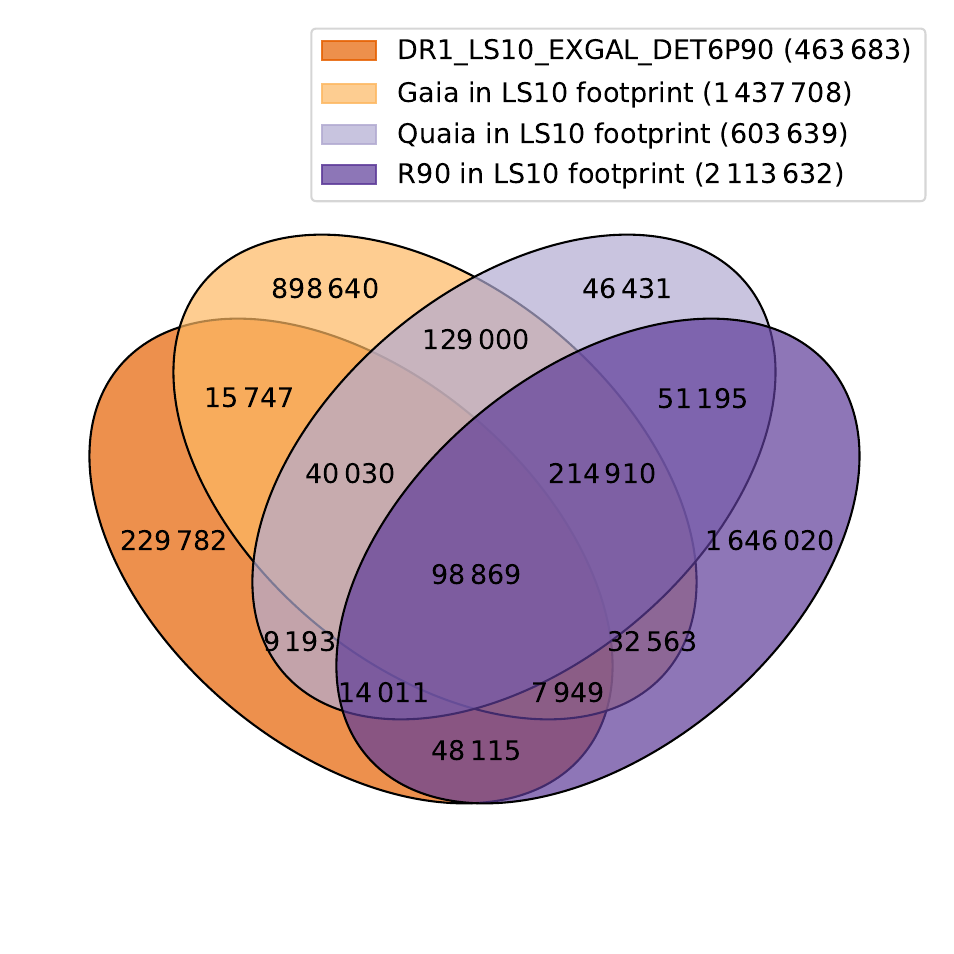}
\caption{Venn diagram distribution for the AGN selected via eROSITA and LS10 (Sample 1), GDR3 high-reliability QSOs, AllWise/R90 and Quaia within the footprint of LS10. In the legend, the number of sources in a given sample, within the footprint of LS10, is reported. In all cases, at least 50\% of the candidate AGN are unique to that selection.}
\label{fig:Venn}
\end{figure}

\noindent Figure \ref{fig:Venn} shows the Venn diagram of the AGN distribution in the four catalogues. Despite the large size of the samples analysed, the overlap remains minimal, with only 98,773 sources common to all selections. It is interesting then to understand what are the properties of the sources in common and the properties of the about 50\% of the eROSITA DR1 sources that are unique detections\footnote{A similar analysis can be done for the  2/3 of the sources that are unique to the AllWISE and Gaia, respectively, but it not the focus of this paper.}. For that, it is important to keep in mind that the Gaia selection, by construction, retains only sources that are best fit by a Gaussian profile, so preferentially point sources or very compact objects. Thus, there is a strong bias toward QSO or the core of local Seyfert 1 galaxies, with a strong contrast between the active nucleus and the host. The bias is retained in the Quaia selection, which requires a detection in unWISE in addition to WISE. Neither the Gaia QSO sample nor Quaia will be able to identify galaxies hosting a low luminosity AGN or at low redshift. In addition, given the shallowness of the data, faint QSOs will not be detected. The AllWISE selection of AGN can also detect sources at lower redshift (regardless of the morphology) or faint sources at higher redshift, but only if dust is heated up sufficiently by the nucleus, assuming dust is actually present. Thus, AllWISE/R90 will not contain low luminosity AGN nor those hosted in early-type galaxies, as they lack dust to be heated up.

\begin{figure*}[!ht]
\centering
\includegraphics[width=\textwidth]{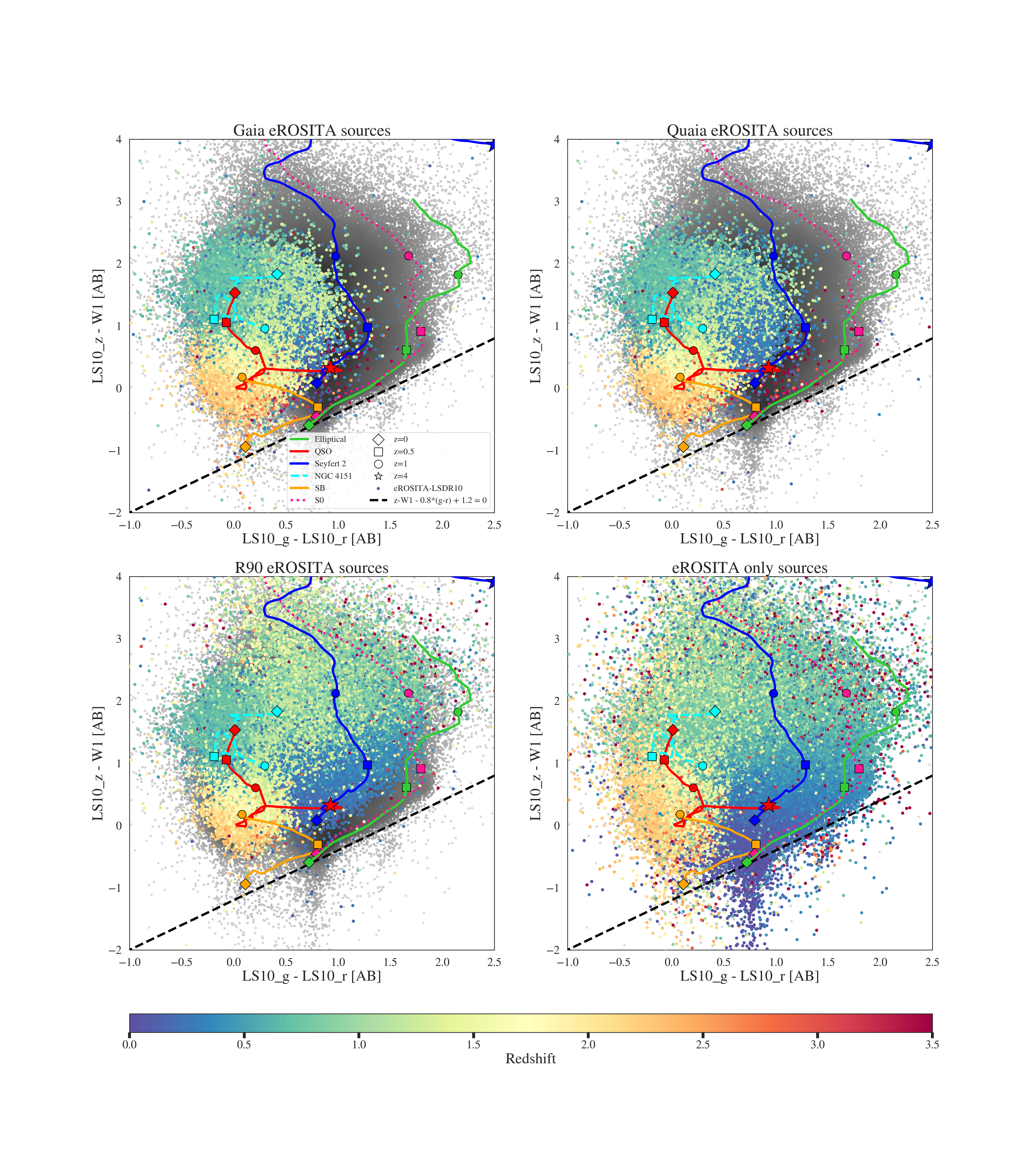}
\caption{{\it g-r} vs {\it z}-W1 colour-colour plots for the eROSITA sample DR1\_LS10\_EXGAL\_DET6P90 eROSITA DR1 sources (grey), detected also in Gaia (top left), Quaia (right left), and AllWISE/R90 (bottom left), respectively. The bottom right panel shows the distribution of the sources that are unique to eROSITA DR1. The sources in all panels are colour-coded as a function of redshift. Overlapped are the tracks of a few examples of typical SEDs, which show that many of the sources that are detected only by eROSITA are early type and Seyfert 2 at redshift below 1.}
\label{fig:AGN_grzw1}
\end{figure*}

\begin{figure}
\centering
\includegraphics[width=0.5\textwidth]{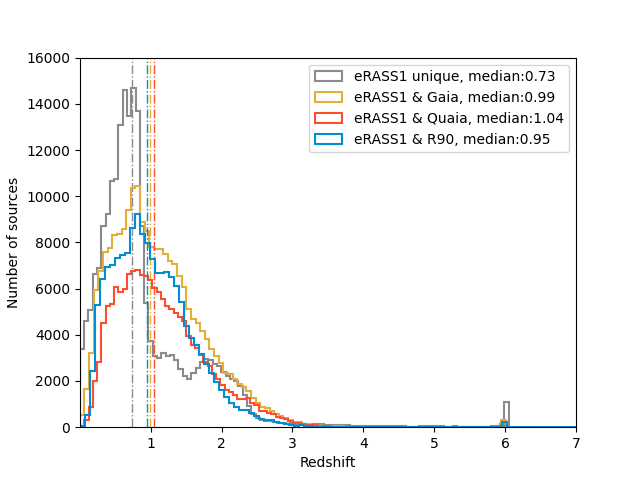}
\caption{ Redshift distribution of the eROSITA/DR1 extragalactic sources in the footprint of LS10, split among the sources that are also detected in Gaia, Quaia, and AllWISE/R90, respectively. The majority of the sources that are unique to eRASS1 are at low redshift, where the contribution from the host galaxy dominates the SED in optical and NIR.}
\label{fig:zhisto}
\end{figure}

All this is confirmed in Figure~\ref{fig:AGN_grzw1} and Figure~\ref{fig:zhisto}.
Figure~\ref{fig:AGN_grzw1} shows the eROSITA/DR1 extragalactic sources that are either detected also by other surveys, or unique, plotted on the basis of they {\it rgzW1} colors, as in \citetalias{Salvato22}, together with the color-color tracks of a few basic SEDs \citep[from][]{Polletta2007, Salvato09}, as a function of redshift. The dominance of QSOs  and spiral galaxies among the sources in common between eROSITA and the other surveys is clear. The additional QSOs that only eROSITA detects will soon be identified by surveys as LSST \citep{Ivezic19}, using faint photometry or variability \citep[LSST, ZTF QSO sample;][]{Nakoneczny2025}.  
Instead, the early-type galaxies hosting an AGN (red and green tracks in the figure) will remain the prerogative of X-ray surveys also in the future, although it will not be surprising if a fraction of the sources will also be detected with the currently ongoing radio surveys \citep[see e.g.,][for the comparison between LOFAR and eROSITA/eFEDS]{Igo2024, Bulbul2022}. 
Figure~\ref{fig:zhisto} shows the histogram of the redshift distribution of the extragalactic eROSITA DR1 sources, split between those detected also by Gaia, Quaia, and AllWISE/R90 or unique to eROSITA/DR1. As discussed above, more than half of the eROSITA only detected sources (65\%), are indeed at redshift z$<$1.

\section{Catalogues format and data access}
\label{Section: DataRelease}
With this paper, we release the three catalogues of counterparts from LS10, GDR3, and CW2020 for the point sources in the Main catalogue of eRASS1 \citep{Merloni24}. They will be made available via Vizier and at the eROSITA website\footnote{\url{https://erosita.mpe.mpg.de/dr1/AllSkySurveyData_dr1/Catalogues_dr1/}} (files 4-12), together with the data model, and the description of the columns. The catalogues include basic columns from the X-ray catalogue and all the relevant columns from the supporting optical/IR catalogues, together with the basic information from {\sc NWAY} like \texttt{p\_any}, \texttt{p\_i}, etc.
 In addition, we also provide  \texttt{p\_any} threshold[6,7,8] at which the completeness and purity curves cross, thus being at the maximum value in that specific eROSITA tile, for a specific \texttt{DET\_LIKE\_0}, that value being \texttt{compur}[6,7,8]. We also provide the value of completeness and purity (completeness[6,7.8] and purity[6,7,8]) corresponding to the \texttt{p\_any} of the source, computed using all the sources in that specific eROSITA tile at the specific \texttt{DET\_LIKE\_0}. In this way, we enable the user to create a sample of sources with a specific completeness and purity over the entire sky. For example, A user could be interested in selecting all the sources with \texttt{DET\_LIKE\_0}$>$7 and purity7$>$0.9.
 This possibility is relevant for studies in which complete control over biases and selection effects is relevant, like in, for example, luminosity function studies. 
A Jupyter notebook is also provided via Zenodo, with instructions to recreate samples 1-6 following the flow chart of Figure \ref{fig:FlowChart}, which can be modified and used to create any other sample.

The priors for the determination of the counterparts, their classification in STAREX, and the model used for the photometric redshifts, as followed in \citet{Saxena2024}, are based on a well-constructed set of training samples, which are also released with this paper. The constructions of these six catalogues  are described in the Appendix \ref{appendix:4XMMBright}. In short, the training samples are from 4XMM \citep{Webb2020} and CSC2 \citep{Evans2024} that have counterparts in LS10, GDR3, and CW2020 (label 1), together with the field sources, i.e. the LS10/GDR3/CW2020 sources that are within 60\,\arcsec\ from the 4XMM/CSC2 position and that are not associated to the X-ray emission (label 0).

\section{Conclusions}\label{section:conclusions}
In this paper, we describe the construction of the three catalogues listing the most reliable counterparts to the point sources detected by eROSITA (PSF Half Energy Width of $\sim$30\arcsec\, and corresponding 1$\sigma$ median positional uncertainty of $\sim$4.5\arcsec and a uniform flux limit at 50\% completeness of $F_{\rm 0.5-2 keV} > 5\times 10^{-14}$ erg s$^{-1}$ cm$^{-2}$) in the first all-sky pass over the Western hemisphere (eROSITA\_DE), using LS10, Gaia DR3, and CW2020, respectively. Each ancillary catalogue (see Section~\ref{section:ancillary}) offers specific advantages and limitations depending on its depth, number of photometric bands, and wavelength coverage. Counterpart identification (Section~\ref{section:counterparts}) was performed using the \texttt{NWAY} algorithm, enhanced with a prior constructed from a Random Forest classifier trained to estimate the probability that a given source is an X-ray emitter based on its SED. The training sample consisted of securely identified X-ray emitters from XMM and Chandra observations. The most reliable counterparts are found within the LS10 footprint, given the depth and the broad multi-band coverage, which provides a rich set of features for defining the SED. Nevertheless, by applying appropriate probability thresholds, the Gaia and CW2020-based matches also yield reliable counterparts, especially in the regions not  covered by LS10.

After identifying the counterparts, a source needs to be classified as Galactic or extragalactic (see Section \ref{section:CTPproperties}). For Gaia, we adopt the classification probabilities (galaxy, star, quasar) provided by the Gaia collaboration, derived from a combination of five methods \citep{Delchambre2023}. These probabilities allow users to apply their own selection criteria based on appropriate cuts. For CW2020, we use a simple cut in W1$-$W2 colour, calibrated by comparing the colour distributions of confirmed extragalactic and Galactic sources. For LS10, we have considered a combination of methods. First, we identify likely stellar sources based on Gaia detections with high proper motion or parallax and point-like morphology.  Those are considered of stellar origin. Then, a machine learning based algorithm, STAREX, was developed within the eROSITA collaboration. This algorithm  combines LS10 photometry and X-ray flux information to assign a probability of the source being extragalactic, using a training set composed of securely identified Galactic and extragalactic X-ray emitters. Finally, for the sources with STAREX extragalactic probabilities below  50\,\%, apply additional diagnostics based on colour-colour and magnitude–X-ray flux diagrams, following the methods described in \citet{Salvato18a, Salvato22}, as well as morphological classification from LS10.. Following this approach, we classify 86,836 sources in the LS10 footprint as stars and 569,778 as extragalactic.

As the final step, we assign redshifts to the extragalactic sources (see Sec.~\ref {sec:redshifts}). First, we compute the photometric redshifts for all extragalactic sources detected in LS10, using the {\sc Circlez} code \citep{Saxena2024}. In addition, we matched the sources  with a compilation of publicly available spectroscopic redshifts, allowing us to asses the quality of our  photometric redshifts. In regions with high-quality photometry across all bands, we achieve an accuracy of 0.067 and an outlier fraction of 12.7\,\%. Outside these areas, the corresponding values are 0.078 and 15\,\%, respectively. For the eRASS1\_GDR3 we provide the redshift from Gaia or Quaia, if available. We also cross-matched the sample the redshift compilation and with our photometric redshift. For the eRASS1\_CW2020 sample, we cross-matched the sources with the same compilation and with the computed photometric redshifts. In all cases, the user may request from the authors the redshift probability distribution function (PDZ) as computed from {\sc Circlez}.

All three  catalogues enable the construction of well-defined samples of X-ray selected sources over the entire sky, which can be compared to samples of similar sources selected using different techniques or at other wavelengths. In the paper (see Section \ref{sec:eRASS1_AGN}), we demonstrate how to construct six AGN samples with varying levels of completeness and purity, both over the extragalactic sky and within the Galactic plane. Focusing on a sample of 463,683 AGN within the LS10 footprint, with an estimated purity of 90\%, we then compare its overlap with three other AGN samples; a Gaia-selected sample, which preferentially  identifies the cores of Seyfert 1 galaxies and QSOs; a sample combining Gaia and unWISE photometry, spectroscopically confirmed via Gaia spectroscopy; a sample  based on AllWISE photometry, with a purity of 90\%.

In the three cases above, purity refers to the reliability of the sources classified as AGN. However, for AGN detected in eRASS1 there is 14\,\% probability of being a spurious X-ray detection. We showed that the counterparts to spurious detections typically have very low p\_any, comparable to those found in  random catalogues.. Applying a threshold of at \texttt{purity6}>0.9 effectively eliminates such spurious detections. The overlap among the four catalogues is marginal, confirming once more that a complete census of AGN requires the combination of multiple selection techniques. Many sources falling in the {\it grz}W1 locus of QSO are not detected by Gaia due to its limited depth. Meanwhile, a significant fraction of AGN detected only by eROSITA with LS10 counterparts are characterised by being at low redshift, with the emission diluted/hidden by the emission of the host galaxy, or they reside in galaxies with minimal dust reprocessing, which could reveal the presence of the AGN in the mid-infrared.

All catalogues constructed in this paper are released to the public (see Section \ref{Section: DataRelease}) via Vizier and the eROSITA web pages at \url{https://erosita.mpe.mpg.de/dr1/AllSkySurveyData_dr1/Catalogues_dr1/} (files 4-12).

\bibliographystyle{aa} 
\bibliography{bibliography.bib}
\newpage
\begin{appendix}

\section{Construction of a reference sample of X-ray sources having secure optical/IR counterparts} 
\label{appendix:4XMMBright}

In this section, we describe the construction of a sample of point-like
X-ray sources having secure optical/IR counterparts from the DESI Legacy
Imaging Surveys catalogue \citep{Dey2019}, the Gaia EDR3 main source catalogue \citep{GaiaEDR3}, and the CatWISE2020 catalogue \citep{Marocco2021}. This reference sample is designed to span
the X-ray flux range probed by the eRASS1 \citep{Merloni24} and eFEDS \citep{Brunner2022}
point source catalogues, and is used to train and validate our cross-matching
algorithm. The selection criteria are tuned
to give a high-purity sample, with completeness sacrificed when necessary.

\subsection{4XMM sample}
We start with the 4XMM DR11 catalogue of detections \citep{Webb2020}, and retain only
pointlike entries (\texttt{SC\_EXT\_ML}$<$4.0 and
\texttt{SC\_EXTENT}$<$3.0\arcsec), having a soft band flux
(0.5--2\,keV, computed as the sum of the native 0.5--1 and 1--2\,keV
bands) greater than $2\times10^{-15}$\,erg\,s$^{-1}$\,cm$^{-2}$, and
with a signal-to-noise ratio on that flux measurement greater than
10. We then apply quality cuts on detection likelihood
(\texttt{SC\_DET\_ML}$>$10.0) and positional precision
(\texttt{SC\_POSERR}$<$1.5\arcsec), and keep only isolated sources
(\texttt{DIST\_NN}$>$10.0\arcsec and \texttt{CONFUSED}=False), that
were observed with low background (\texttt{HIGH\_BACKGROUND}=False)
with reasonable exposure time (\texttt{EP\_ONTIME}$>$5\,ks),
are not flagged (\texttt{SC\_SUM\_FLAG}$\le$15), and which were not
too far-off axis ($<$15\arcmin\ in at least one instrument).  Finally, we apply spatial filtering, removing any detections that lie at low
Galactic latitudes ($|b|<10$\,deg), near the centres of the Large
Magellanic Cloud (5\,deg radius), the Small Magellanic Cloud (3\,deg
radius) or M31 (1\,deg radius), within 0.05\,deg of stars from the Yale
Bright Star catalogue \citep{Hoffleit1991cat}, within 0.05\,deg of stars in
Tycho-2 \citep{Hog2000} having $VT<9$\,mag, or within the disc (radius derived from $D25$) of nearby bright
($B\_T<12$\,mag) galaxies from the RC3 catalogue \citep{deVacoulers1991}.  After retaining
a single detection per unique X-ray source, we are left with a clean sample of 47\,973
4XMM sources.

\subsection{CSC2 sample}
We retrieved a filtered sample of sources from the CSC2.0
\texttt{primary\_source} table via the database query web
API\footnote{\url{https://cxc.cfa.harvard.edu/csc/cli/index.html}}.
We selected bright ACIS detections with 0.5--2\,keV aperture fluxes
$>2\times10^{-15}$\,erg\,s$^{-1}$\,cm$^{-2}$, having a detection
\texttt{significance}$>$6, that were point-like
(\texttt{major\_axis\_b}$<$1\,arcsec), well localised
(\texttt{err\_ellipse\_r0}$<$1\,arcsec), observed with reasonable
exposure time ($>$1\,ks), and those that were not flagged as piled up,
confused, or associated with a read-out streak. As before, we excluded
any sources which lie at low Galactic latitudes, near the LMC, SMC or M31, or those that were close to very bright stars or bright galaxies.
After these filtering steps, our CSC2.0 sample contains 7426 sources.

\subsection{Association with optical/IR counterpart catalogues}\label{app: ctp_to_training}

In order to provide samples of clean X-ray sources with secure
optical/IR counterparts, we carried out the following procedure to match each X-ray reference sample with the DESI Legacy Imaging Survey  DR10\footnote{We used an early internally released version of LS DR10, which is identical to the public version in $>90$\,\% of survey bricks.}, Gaia
EDR3 and CatWISE2020 catalogues. The following was repeated for each
combination of X-ray and optical/IR catalogue.  Firstly, we retrieved
an initial catalogue of all objects in the optical/IR catalogue within a radius of 60\arcsec\ of the X-ray positions - these initial catalogues serve the dual purposes of i) being the parent sample from
which we find optical/IR counterparts, and ii) providing a pool of
optical/IR objects which we can be certain are {\em not} strong X-ray
emitters. Given the $\sim$20-year baselines spanned by the X-ray
samples, we used proper motion estimates in the optical/IR catalogues
to propagate the coordinates to the epoch of the nearest X-ray
detection.  Then, using only spatial information (positions,
uncertainties), the \texttt{NWAY} tool \citep{Salvato18b} was used to find the best
optical/IR counterpart (\texttt{match\_flag}=1) for each X-ray source.
Since we are interested only in a highly pure sample, we further filtered the samples to retain only secure cross-matches where
\texttt{p\_any}$>$0.3 and \texttt{p\_i}$>$0.5.  As a further refinement, 
an even higher quality
sub-sample was identified, consisting of cross-matches having
\texttt{p\_any}$>$0.9 and \texttt{p\_i}$>$0.8. The number of objects in our training samples is given in Table \ref{tab:a1_training_sample}.

\begin{table}
    \centering
    \begin{tabular}{ccccc}
      O/IR Catalogue    & $N_\mathrm{4XMM}$ & $N_\mathrm{CSC2}$  & \\
      \hline
      LS10    &  39636 (32900)  & 5888 (5643) \\
      Gaia EDR3    &  27219 (26129)  & 3477 (3379) \\
      CatWISE2020  &  44218 (40399)  & 6204 (5707) \\
    \end{tabular}
    \caption{Number of X-ray training sample sources (from 4XMM and CSC2) 
    having secure counterparts in each
    of the three optical/IR catalogues considered here. The numbers in brackets are for the highest quality sub-sample.}
    \label{tab:a1_training_sample}
\end{table}

To inform the cross-matching algorithm presented in this work, we 
generated a paired `field catalogue' (i.e., non-X-ray emitting optical/IR objects) for each combination of X-ray and optical/IR catalogue. Each field catalogue is made up of all the optical/IR sources within an annulus (inner radius 15\arcsec, outer 60\arcsec) around the X-ray source positions, further excluding 15\arcsec\ radius regions near {\em any} other X-ray sources found in the full 4XMM DR11 catalogue, or the CSC2 catalogue (considering only significantly detected compact sources with 0.5--2\,keV aperture fluxes $>5\times10^{-16}$\,erg\,s$^{-1}$\,cm$^{-2}$). This approach ensures that the field catalogue probes a very similar distribution of survey conditions (both astrophysical and those associated with spatial variations in data quality), as the X-ray reference samples themselves.

\begin{figure*}[!t]
    \centering
    \includegraphics[width=\textwidth]{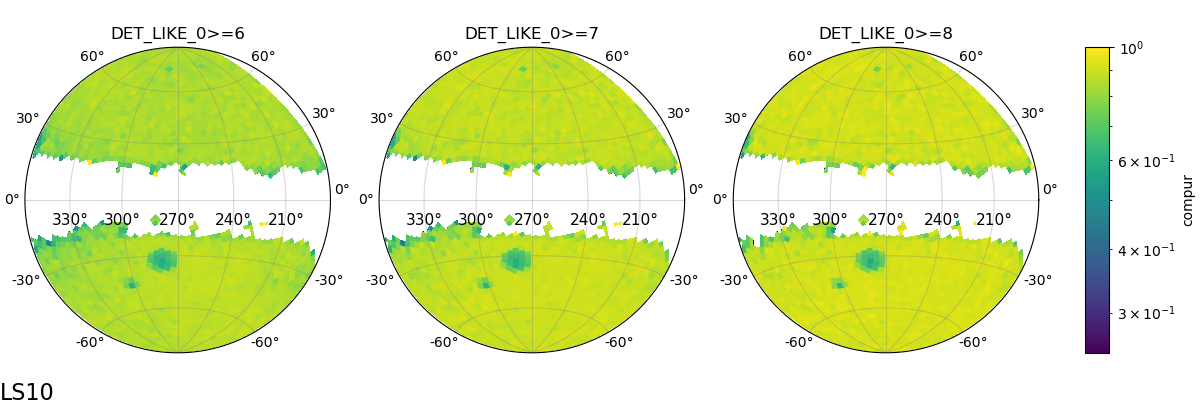}
\includegraphics[width=\textwidth]{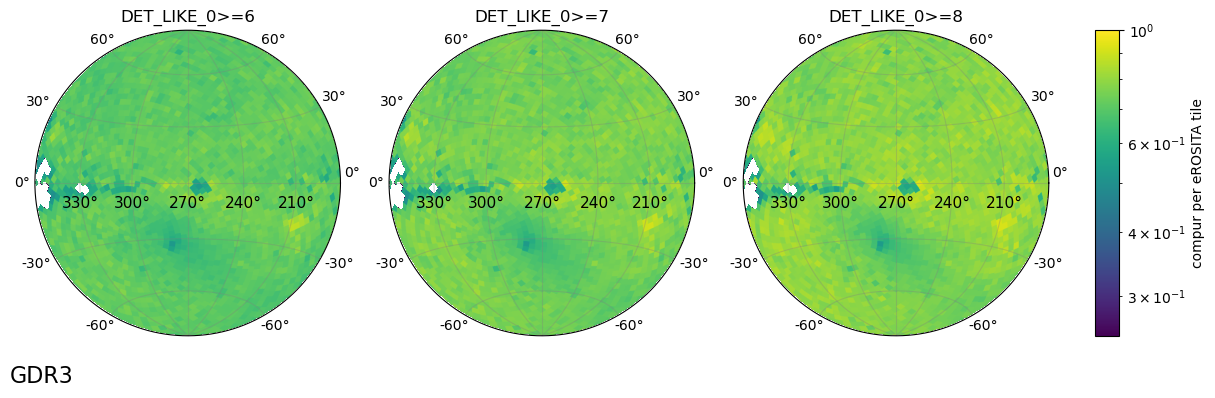}
\includegraphics[width=\textwidth]{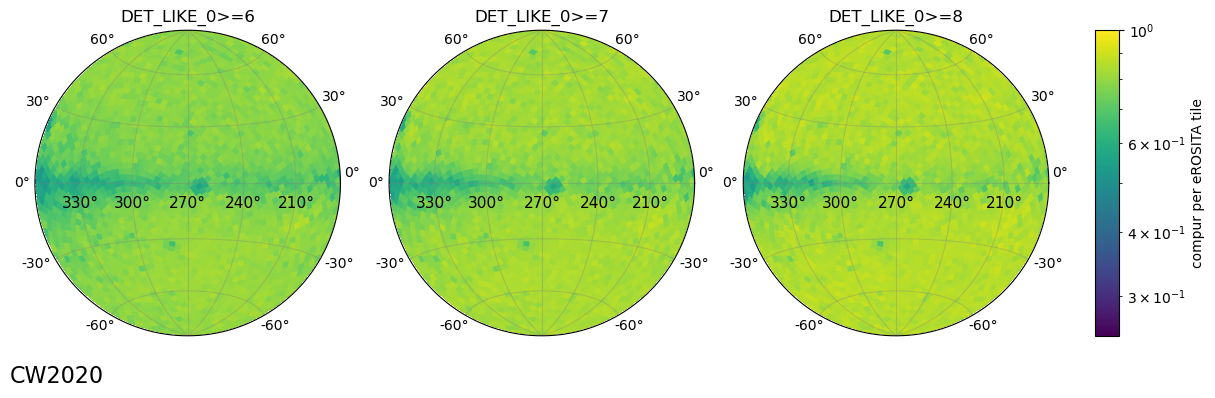}
    \caption{Map of the mean value per eROSITA tile of completeness and purity (\texttt{compur[6,7,8]} in the catalogues) at the intersection point (see Figure \ref{fig:Meanpurcomp} for a visualization) as a function of DET\_LIKE\_0, for LS10 (left panel), GDR3 (middle panel), and CW2020 (right panel). With the exception of the Magellanic Clouds and the Galactic Plane, completeness and purity are high, and they increase with the detection likelihood. There are a few tiles in the Gaia maps where the completeness and purity curves do not intersect due to the limited number of sources.}
    \label{fig:compurMap}
\end{figure*}

\section {More examples of construction of AGN samples within the footprint of LS10 }\label{appendix:LS10AGNSample}
In Section \ref{subsec:DR1LS10} we described the construction of a highly pure (purity>90\%) sample of eRASS1 AGN with counterparts in LS10. However, that selection included sources with different optical properties, depending on the depth of the LS10 data available in that position. A user could be interested in generating a sample with similar optical data. For that, samples 2,3, and 4 in Figure~\ref{fig:FlowChart} would be more suitable.
The first steps are the same as for Sample 1, but then we would consider the sources that are within the area with all( at least one) optical bands with nominal depth (\texttt{inAllLS10}$=$1/\texttt{inAnyLS10}$=$1). To preserve high completeness, we can choose to select the sources that have \texttt{DET\_LIKE\_0}$\ge$6 (Sample 2). As reported in \citet{Seppi2022} at this threshold, we expect up to 14\% spurious detections, but the fraction is highly reduced once only sources with p\_any above the threshold (see Section \ref{section:counterparts}) are considered. However, the fraction of spurious X-ray detections could also be limited by selecting a higher detection likelihood (\texttt{DET\_LIKE\_0}$\ge$7 or \texttt{DET\_LIKE\_0}$\ge$8) and  limiting the sources to those with p\_any above the threshold that simultaneously maximise completeness and purity at that detection likelihood.

\section{Acknowledgments}
    We are gratefull to the referee who read the manuscript carefully and provided helpful suggestions. 
    AG acknowledges funding from the Hellenic Foundation for Research and Innovation (HFRI) project "4MOVE-U" grant agreement 2688, which is part of the programme "2nd Call for HFRI Research Projects to support Faculty Members and Researchers".
    Ms and WR acknowledge the support (Förderkennzeichen 50002207) from Deutsches Zentrum für Luft- und Raumfahrt (DLR). ZI, JW, AM, and MS acknowledge the support by the Excellence Cluster ORIGINS, which is funded by the Deutsche Forschungsgemeinschaft (DFG, German Research Foundation) under Germany´s Excellence Strategy – EXC-2094 – 390783311.
    IM acknowledges financial support from the Severo Ochoa grant CEX2021-001131-S funded by MCIN/AEI/10.13039/501100011033 and  by the Spanish Ministry of Science, Innovation y University (MCIU) under the grant PID2022-140871NB-C21.
    JR acknowledges support from DLR under 50QR2505.
    This work is based on data from eROSITA, the soft X-ray instrument aboard SRG, a joint Russian-German science mission supported by the Russian Space Agency (Roskosmos), in the interests of the Russian Academy of Sciences represented by its Space Research Institute (IKI), and the Deutsches Zentrum für Luft- und Raumfahrt (DLR). The SRG spacecraft was built by Lavochkin Association (NPOL) and its subcontractors, and is operated by NPOL with support from the Max Planck Institute for Extraterrestrial Physics (MPE).
     \newline
    The development and construction of the eROSITA X-ray instrument was led by MPE, with contributions from the Dr. Karl Remeis Observatory Bamberg \& ECAP (FAU Erlangen-Nuernberg), the University of Hamburg Observatory, the Leibniz Institute for Astrophysics Potsdam (AIP), and the Institute for Astronomy and Astrophysics of the University of Tübingen, with the support of DLR and the Max Planck Society. The Argelander Institute for Astronomy of the University of Bonn and the Ludwig Maximilians Universität Munich also participated in the science preparation for eROSITA.
     \newline
    We have made use of TOPCAT and STILTS \citep[][]{Taylor05, Taylor06}, scipy, astropy, scikit-learn, matplotlib.
    \newline
    Based on observations made with ESO Telescopes at the La Silla Paranal Observatory under programme IDs 177.A-3016, 177.A-3017, 177.A-3018 and 179.A-2004, and on data products produced by the KiDS consortium. The KiDS production team acknowledges support from: Deutsche Forschungsgemeinschaft, ERC, NOVA and NWO-M grants; Target; the University of Padova, and the University Federico II (Naples).
    \newline
The data presented here were obtained in part with ALFOSC, which is provided by the Instituto de Astrofísica de Andalucía (IAA-CSIC) under a joint agreement with the University of Copenhagen and NOT. 
Some of the observations were collected at the Centro Astronómico Hispano Alemán (CAHA) at Calar Alto, operated jointly by the
Instituto de Astrofísica de Andalucía (IAA-CSIC) and Junta de Andalucía.
\newline
The Legacy Surveys consist of three individual and complementary projects: the Dark Energy Camera Legacy Survey (DECaLS; Proposal ID 2014B-0404; PIs: David Schlegel and Arjun Dey), the Beijing-Arizona Sky Survey (BASS; NOAO Prop. ID 2015A-0801; PIs: Zhou Xu and Xiaohui Fan), and the Mayall z-band Legacy Survey (MzLS; Prop. ID 2016A-0453; PI: Arjun Dey). DECaLS, BASS and MzLS together include data obtained, respectively, at the Blanco telescope, Cerro Tololo Inter-American Observatory, NSF’s NOIRLab; the Bok telescope, Steward Observatory, University of Arizona; and the Mayall telescope, Kitt Peak National Observatory, NOIRLab. The Legacy Surveys project is honored to be permitted to conduct astronomical research on Iolkam D\'uag (Kitt Peak), a mountain with particular significance to the Tohono O’odham Nation.
\newline
This research used data obtained with the Dark Energy Spectroscopic Instrument (DESI). DESI construction and operations is managed by the Lawrence Berkeley National Laboratory. This material is based upon work supported by the U.S. Department of Energy, Office of Science, Office of High-Energy Physics, under Contract No. DE–AC02–05CH11231, and by the National Energy Research Scientific Computing Center, a DOE Office of Science User Facility under the same contract. Additional support for DESI was provided by the U.S. National Science Foundation (NSF), Division of Astronomical Sciences under Contract No. AST-0950945 to the NSF’s National Optical-Infrared Astronomy Research Laboratory; the Science and Technology Facilities Council of the United Kingdom; the Gordon and Betty Moore Foundation; the Heising-Simons Foundation; the French Alternative Energies and Atomic Energy Commission (CEA); the National Council of Humanities, Science and Technology of Mexico (CONAHCYT); the Ministry of Science and Innovation of Spain (MICINN), and by the DESI Member Institutions: www.desi.lbl.gov/collaborating-institutions. The DESI collaboration is honored to be permitted to conduct scientific research on I’oligam Du’ag (Kitt Peak), a mountain with particular significance to the Tohono O’odham Nation. Any opinions, findings, and conclusions or recommendations expressed in this material are those of the author(s) and do not necessarily reflect the views of the U.S. National Science Foundation, the U.S. Department of Energy, or any of the listed funding agencies.
\newline
This work has made use of data from the European Space Agency (ESA) mission
{\it Gaia} (\url{https://www.cosmos.esa.int/gaia}), processed by the {\it Gaia}
Data Processing and Analysis Consortium (DPAC,
\url{https://www.cosmos.esa.int/web/gaia/dpac/consortium}). Funding for the DPAC
has been provided by national institutions, in particular the institutions
participating in the {\it Gaia} Multilateral Agreement.
\newline
Funding for the Sloan Digital Sky Survey V has been provided by the Alfred P. Sloan Foundation, the Heising-Simons Foundation, the National Science Foundation, and the Participating Institutions. SDSS acknowledges support and resources from the Center for High-Performance Computing at the University of Utah. SDSS telescopes are located at Apache Point Observatory, funded by the Astrophysical Research Consortium and operated by New Mexico State University, and at Las Campanas Observatory, operated by the Carnegie Institution for Science. The SDSS website is \url{www.sdss.org}.

SDSS is managed by the Astrophysical Research Consortium for the Participating Institutions of the SDSS Collaboration, including Caltech, The Carnegie Institution for Science, Chilean National Time Allocation Committee (CNTAC) ratified researchers, The Flatiron Institute, the Gotham Participation Group, Harvard University, Heidelberg University, The Johns Hopkins University, L'Ecole polytechnique f\'{e}d\'{e}rale de Lausanne (EPFL), Leibniz-Institut f\"{u}r Astrophysik Potsdam (AIP), Max-Planck-Institut f\"{u}r Astronomie (MPIA Heidelberg), Max-Planck-Institut f\"{u}r Extraterrestrische Physik (MPE), Nanjing University, National Astronomical Observatories of China (NAOC), New Mexico State University, The Ohio State University, Pennsylvania State University, Smithsonian Astrophysical Observatory, Space Telescope Science Institute (STScI), the Stellar Astrophysics Participation Group, Universidad Nacional Aut\'{o}noma de M\'{e}xico, University of Arizona, University of Colorado Boulder, University of Illinois at Urbana-Champaign, University of Toronto, University of Utah, University of Virginia, Yale University, and Yunnan University.
\newline
This paper uses observations made at the South African Astronomical Observatory (SAAO)

\end{appendix}
\end{document}